\documentclass[twocolumn,letterpaper,aps,prc,superscriptaddress,showpacs,nofootinbib,floatfix]{revtex4}

\usepackage{graphicx}
\usepackage{multirow}

\begin{document}


\title{Deviation from quark number scaling of the anisotropy parameter $v_2$ of
pions, kaons, and protons in Au+Au collisions at $\sqrt{s_{NN}}$ = 200 GeV}

\newcommand{\abilene}{Abilene Christian University, Abilene, Texas 79699, USA}
\newcommand{\banaras}{Department of Physics, Banaras Hindu University, Varanasi 221005, India}
\newcommand{\barc}{Bhabha Atomic Research Centre, Bombay 400 085, India}
\newcommand{\bnlcoll}{Collider-Accelerator Department, Brookhaven National Laboratory, Upton, New York 11973-5000, USA}
\newcommand{\bnlphys}{Physics Department, Brookhaven National Laboratory, Upton, New York 11973-5000, USA}
\newcommand{\caucr}{University of California - Riverside, Riverside, California 92521, USA}
\newcommand{\charlesczech}{Charles University, Ovocn\'{y} trh 5, Praha 1, 116 36, Prague, Czech Republic}
\newcommand{\chonbuk}{Chonbuk National University, Jeonju, 561-756, Korea}
\newcommand{\ciae}{Science and Technology on Nuclear Data Laboratory, China Institute of Atomic Energy, Beijing 102413, P.~R.~China}
\newcommand{\cns}{Center for Nuclear Study, Graduate School of Science, University of Tokyo, 7-3-1 Hongo, Bunkyo, Tokyo 113-0033, Japan}
\newcommand{\colorado}{University of Colorado, Boulder, Colorado 80309, USA}
\newcommand{\columbia}{Columbia University, New York, New York 10027 and Nevis Laboratories, Irvington, New York 10533, USA}
\newcommand{\czechtech}{Czech Technical University, Zikova 4, 166 36 Prague 6, Czech Republic}
\newcommand{\dapnia}{Dapnia, CEA Saclay, F-91191, Gif-sur-Yvette, France}
\newcommand{\debrecen}{Debrecen University, H-4010 Debrecen, Egyetem t{\'e}r 1, Hungary}
\newcommand{\elte}{ELTE, E{\"o}tv{\"o}s Lor{\'a}nd University, H - 1117 Budapest, P{\'a}zm{\'a}ny P. s. 1/A, Hungary}
\newcommand{\ewha}{Ewha Womans University, Seoul 120-750, Korea}
\newcommand{\fit}{Florida Institute of Technology, Melbourne, Florida 32901, USA}
\newcommand{\fsu}{Florida State University, Tallahassee, Florida 32306, USA}
\newcommand{\gsu}{Georgia State University, Atlanta, Georgia 30303, USA}
\newcommand{\hiroshima}{Hiroshima University, Kagamiyama, Higashi-Hiroshima 739-8526, Japan}
\newcommand{\ihepprot}{IHEP Protvino, State Research Center of Russian Federation, Institute for High Energy Physics, Protvino, 142281, Russia}
\newcommand{\illuiuc}{University of Illinois at Urbana-Champaign, Urbana, Illinois 61801, USA}
\newcommand{\inrras}{Institute for Nuclear Research of the Russian Academy of Sciences, prospekt 60-letiya Oktyabrya 7a, Moscow 117312, Russia}
\newcommand{\instpasczech}{Institute of Physics, Academy of Sciences of the Czech Republic, Na Slovance 2, 182 21 Prague 8, Czech Republic}
\newcommand{\isu}{Iowa State University, Ames, Iowa 50011, USA}
\newcommand{\jinrdubna}{Joint Institute for Nuclear Research, 141980 Dubna, Moscow Region, Russia}
\newcommand{\jyvaskyla}{Helsinki Institute of Physics and University of Jyv{\"a}skyl{\"a}, P.O.Box 35, FI-40014 Jyv{\"a}skyl{\"a}, Finland}
\newcommand{\kek}{KEK, High Energy Accelerator Research Organization, Tsukuba, Ibaraki 305-0801, Japan}
\newcommand{\korea}{Korea University, Seoul, 136-701, Korea}
\newcommand{\kurchatov}{Russian Research Center ``Kurchatov Institute", Moscow, 123098 Russia}
\newcommand{\kyoto}{Kyoto University, Kyoto 606-8502, Japan}
\newcommand{\labllr}{Laboratoire Leprince-Ringuet, Ecole Polytechnique, CNRS-IN2P3, Route de Saclay, F-91128, Palaiseau, France}
\newcommand{\lawllnl}{Lawrence Livermore National Laboratory, Livermore, California 94550, USA}
\newcommand{\losalamos}{Los Alamos National Laboratory, Los Alamos, New Mexico 87545, USA}
\newcommand{\lpc}{LPC, Universit{\'e} Blaise Pascal, CNRS-IN2P3, Clermont-Fd, 63177 Aubiere Cedex, France}
\newcommand{\lund}{Department of Physics, Lund University, Box 118, SE-221 00 Lund, Sweden}
\newcommand{\maryland}{University of Maryland, College Park, Maryland 20742, USA}
\newcommand{\mass}{Department of Physics, University of Massachusetts, Amherst, Massachusetts 01003-9337, USA }
\newcommand{\muenster}{Institut fur Kernphysik, University of Muenster, D-48149 Muenster, Germany}
\newcommand{\muhlenberg}{Muhlenberg College, Allentown, Pennsylvania 18104-5586, USA}
\newcommand{\myongji}{Myongji University, Yongin, Kyonggido 449-728, Korea}
\newcommand{\nagasaki}{Nagasaki Institute of Applied Science, Nagasaki-shi, Nagasaki 851-0193, Japan}
\newcommand{\newmex}{University of New Mexico, Albuquerque, New Mexico 87131, USA }
\newcommand{\nmsu}{New Mexico State University, Las Cruces, New Mexico 88003, USA}
\newcommand{\ornl}{Oak Ridge National Laboratory, Oak Ridge, Tennessee 37831, USA}
\newcommand{\orsay}{IPN-Orsay, Universite Paris Sud, CNRS-IN2P3, BP1, F-91406, Orsay, France}
\newcommand{\peking}{Peking University, Beijing 100871, P.~R.~China}
\newcommand{\pnpi}{PNPI, Petersburg Nuclear Physics Institute, Gatchina, Leningrad region, 188300, Russia}
\newcommand{\riken}{RIKEN Nishina Center for Accelerator-Based Science, Wako, Saitama 351-0198, Japan}
\newcommand{\rikjrbrc}{RIKEN BNL Research Center, Brookhaven National Laboratory, Upton, New York 11973-5000, USA}
\newcommand{\rikkyo}{Physics Department, Rikkyo University, 3-34-1 Nishi-Ikebukuro, Toshima, Tokyo 171-8501, Japan}
\newcommand{\saispbstu}{Saint Petersburg State Polytechnic University, St. Petersburg, 195251 Russia}
\newcommand{\saopaulo}{Universidade de S{\~a}o Paulo, Instituto de F\'{\i}sica, Caixa Postal 66318, S{\~a}o Paulo CEP05315-970, Brazil}
\newcommand{\seoulnat}{Seoul National University, Seoul, Korea}
\newcommand{\stonybrkc}{Chemistry Department, Stony Brook University, SUNY, Stony Brook, New York 11794-3400, USA}
\newcommand{\stonycrkp}{Department of Physics and Astronomy, Stony Brook University, SUNY, Stony Brook, New York 11794-3400, USA}
\newcommand{\tenn}{University of Tennessee, Knoxville, Tennessee 37996, USA}
\newcommand{\titech}{Department of Physics, Tokyo Institute of Technology, Oh-okayama, Meguro, Tokyo 152-8551, Japan}
\newcommand{\tsukuba}{Institute of Physics, University of Tsukuba, Tsukuba, Ibaraki 305, Japan}
\newcommand{\vandy}{Vanderbilt University, Nashville, Tennessee 37235, USA}
\newcommand{\waseda}{Waseda University, Advanced Research Institute for Science and Engineering, 17 Kikui-cho, Shinjuku-ku, Tokyo 162-0044, Japan}
\newcommand{\weizmann}{Weizmann Institute, Rehovot 76100, Israel}
\newcommand{\wigner}{Institute for Particle and Nuclear Physics, Wigner Research Centre for Physics, Hungarian Academy of Sciences (Wigner RCP, RMKI) H-1525 Budapest 114, POBox 49, Budapest, Hungary}
\newcommand{\yonsei}{Yonsei University, IPAP, Seoul 120-749, Korea}
\affiliation{\abilene}
\affiliation{\banaras}
\affiliation{\barc}
\affiliation{\bnlcoll}
\affiliation{\bnlphys}
\affiliation{\caucr}
\affiliation{\charlesczech}
\affiliation{\chonbuk}
\affiliation{\ciae}
\affiliation{\cns}
\affiliation{\colorado}
\affiliation{\columbia}
\affiliation{\czechtech}
\affiliation{\dapnia}
\affiliation{\debrecen}
\affiliation{\elte}
\affiliation{\ewha}
\affiliation{\fit}
\affiliation{\fsu}
\affiliation{\gsu}
\affiliation{\hiroshima}
\affiliation{\ihepprot}
\affiliation{\illuiuc}
\affiliation{\inrras}
\affiliation{\instpasczech}
\affiliation{\isu}
\affiliation{\jinrdubna}
\affiliation{\jyvaskyla}
\affiliation{\kek}
\affiliation{\korea}
\affiliation{\kurchatov}
\affiliation{\kyoto}
\affiliation{\labllr}
\affiliation{\lawllnl}
\affiliation{\losalamos}
\affiliation{\lpc}
\affiliation{\lund}
\affiliation{\maryland}
\affiliation{\mass}
\affiliation{\muenster}
\affiliation{\muhlenberg}
\affiliation{\myongji}
\affiliation{\nagasaki}
\affiliation{\newmex}
\affiliation{\nmsu}
\affiliation{\ornl}
\affiliation{\orsay}
\affiliation{\peking}
\affiliation{\pnpi}
\affiliation{\riken}
\affiliation{\rikjrbrc}
\affiliation{\rikkyo}
\affiliation{\saispbstu}
\affiliation{\saopaulo}
\affiliation{\seoulnat}
\affiliation{\stonybrkc}
\affiliation{\stonycrkp}
\affiliation{\tenn}
\affiliation{\titech}
\affiliation{\tsukuba}
\affiliation{\vandy}
\affiliation{\waseda}
\affiliation{\weizmann}
\affiliation{\wigner}
\affiliation{\yonsei}
\author{A.~Adare} \affiliation{\colorado}
\author{S.~Afanasiev} \affiliation{\jinrdubna}
\author{C.~Aidala} \affiliation{\mass}
\author{N.N.~Ajitanand} \affiliation{\stonybrkc}
\author{Y.~Akiba} \affiliation{\riken} \affiliation{\rikjrbrc}
\author{H.~Al-Bataineh} \affiliation{\nmsu}
\author{J.~Alexander} \affiliation{\stonybrkc}
\author{K.~Aoki} \affiliation{\kyoto} \affiliation{\riken}
\author{Y.~Aramaki} \affiliation{\cns}
\author{E.T.~Atomssa} \affiliation{\labllr}
% ...
\collaboration{PHENIX Collaboration} \noaffiliation

\date{\today}


\begin{abstract}

Measurements of the anisotropy parameter $v_2$ of identified
hadrons (pions, kaons, and protons) as a function of centrality,
transverse momentum $p_T$, and transverse kinetic energy KE$_T$ at
midrapidity ($|\eta|<0.35$) in Au+Au collisions at
$\sqrt{s_{NN}}$~=~200~GeV are presented.  Pions and protons are
identified up to $p_T=$ 6~GeV/$c$, and kaons up to $p_T=$
4~GeV/$c$, by combining information from time-of-flight and
aerogel \v{C}erenkov detectors in the PHENIX Experiment. The scaling of
$v_2$ with the number of valence quarks ($n_q$) has been studied
in different centrality bins as a function of transverse momentum
and transverse kinetic energy. A deviation from previously observed 
quark-number scaling is observed at large values of KE$_T/n_q$ 
in noncentral Au+Au collisions (20--60\%), but this scaling remains
valid in central collisions (0--10\%).

\end{abstract}


\maketitle

\section{Introduction}

Measurements of the anisotropy parameter $v_2$ (the second
coefficient in the Fourier expansion of the hadron yields with
respect to the reaction plane) have played a pivotal role in the
discovery of the strongly coupled quark-gluon plasma (sQGP) at
RHIC~\cite{Adcox:2004mh,Adams:2005dq,Back:2004je,Arsene:2004fa}.
At low $p_T$ ($\le$~2~GeV/$c$) the agreement between ideal
hydrodynamics calculations and the data have led to the conclusion
that a near-perfect fluid is created in heavy-ion collisions at
RHIC~\cite{Shuryak:2004cy, Gyulassy:2004zy}. Recent theoretical
efforts aiming to quantify the ratio of the shear viscosity to the
entropy density $\eta/s$ (see for example reviews
in~\cite{Heinz:2009xj,Romatschke:2009im,Teaney:2009qa}) have
confirmed that in the sQGP fluid this ratio is close to a
conjectured quantum limit~\cite{Kovtun:2004de}. The high $p_T$
($\ge$~6~GeV/$c$) azimuthal
anisotropies~\cite{Just:2007ju,Afanasiev:2009iv,Adare:2010sp} have
been attributed to the path-length dependence of energy loss in
the medium and are used to constrain the theoretical descriptions
of jet energy loss~\cite{Bass:2008rv,Wang:2001xn}. At intermediate
$p_T$ (2--6~GeV/$c$), which is the focus of this paper, the
identified hadron anisotropies have shown strong evidence for
quark-like degrees of freedom and significant collectivity at the
parton level. This is supported by the observation of scaling with
the number of valence quarks in the hadron ($n_q$
scaling)~\cite{Adler:2003kt,Adams:2003am,Adare:2006ti,Afanasiev:2007tv,Abelev:2007qg,Abelev:2008ed}.

The scaling with number of valence quarks ($n_q$) was seen as a
confirmation of quark recombination as a novel particle-production
mechanism that competes with fragmentation in the 
intermediate-$p_T$ range.  Recombination
models~\cite{Hwa:2002tu,Fries:2003vb,Fries:2003kq,Greco:2003xt,Molnar:2003ff}
were developed to account for the unusually large baryon-to-meson
ratios (relative to $p$+$p$ collisions) and nuclear-modification
factors~\cite{Adler:2003kg,Adler:2003cb,Adams:2003am}, as well as
the large elliptic flow at intermediate $p_T$, with pronounced
differences between baryons and mesons. In the models, the $n_q$
scaling, which is manifested as $v^{hadron}_{2}(p_T) \approx
n_qv_2(p_T/n_q)$, is an approximate scaling that comes from the
addition of the valence-quark momenta at hadronization, with the
assumption that the collective flow develops at the partonic
level.

There are several theoretical considerations that suggest that the
$n_q$ scaling should be violated in certain conditions. For
example, the inclusion of higher Fock states describing the
contribution of sea quarks and gluons have been shown to affect
the $n_q$ scaling~\cite{Muller:2005pv}. Similarly, models that
consider recombination between ``thermal'' partons (soft partons
thermalized in the medium) and ``shower'' partons (partons fragmented
from jets) predict centrality-dependent deviations from $n_q$
scaling that are particle-species dependent~\cite{Hwa:0801aa}.
Understanding the limits of the recombination domain is important
in relation to viscous hydrodynamics and the extraction of the
shear viscosity over entropy density ($\eta/s$) from the
data~\cite{Ferini:2008he,Dusling:2009df,Lacey:2010fe}, as well as
for developing a unified approach in describing jet energy loss
and high $p_T$
$v_2$~\cite{Fochler:2008ts,Lacey:2009ps,PhysRevC.69.034908}.
Searches for deviations from $n_q$ scaling are also important for
the low-energy scan program at RHIC as they have been considered
as a signature of the transition between sQGP formation and a
hadronic system. Recent considerations of baryon transport may
complicate this picture~\cite{Dunlop:2011cf}, which further
reinforces the need for a detailed understanding of this scaling
at $\sqrt{s_{NN}}$ = 200~GeV.

The $n_q$ scaling has been tested in certain centralities and
$p_{T}$ regions with identified
particles~\cite{Adler:2003kt,Adams:2003am,Adare:2006ti,Afanasiev:2007tv,Abelev:2007qg,Abelev:2008ed}.
However, the precision of experimental data on identified hadron
$v_2$ is in many cases limited in statistics and $p_T$ reach,
especially for baryon measurements at KE$_{T}/n_q>$~1~GeV (KE$_T =
\sqrt{p_T^2+m_0^2}-m_0$) where the $n_q$ scaling may start to
break. Therefore, the detailed $p_T$ limits and centrality
dependence of the $n_q$ scaling have not been tested.

This paper reports on high-statistics measurements of the second
order Fourier coefficient $v_2$ for identified pions
($\pi^{+}+\pi^{-}$), kaons ($K^{+}+K^{-}$), and protons
($p+\bar{p}$), which extend to relatively high $p_T$ (up to
6~GeV/$c$ for pions and protons and 4~GeV/$c$ for kaons). The data
for different centrality events (0--10\%, 10--20\%, 20--40\%,
40--60\%, and combinations thereof) are analyzed separately and
the $n_q$ scaling is examined as a function of centrality.
Comparisons with published measurements of $K^{0}_{S}$ and
$\Lambda$ from STAR collaboration~\cite{Abelev:2008ed} are shown
in the centralities 0--10\% and 10--40\%. The experimental details
are presented in Section~\ref{s:exp}, the analysis methods are in
Section~\ref{s:analysis}, the results and discussion are in
Section~\ref{s:results}, and Section~\ref{s:summary} summarizes
our findings.

\section{Experimental Setup}

\label{s:exp} The PHENIX experiment is designed for the study of
nuclear matter in extreme conditions through a variety of
experimental observables. It comprises a tracking system optimized
for the high-multiplicity environment of ultra-relativistic heavy
ion collisions, a set of particle identification (PID) detectors,
and a set of detectors aimed at determining the global properties
of the collisions. 

\begin{figure}[htbp]
\includegraphics[width=0.96\linewidth]{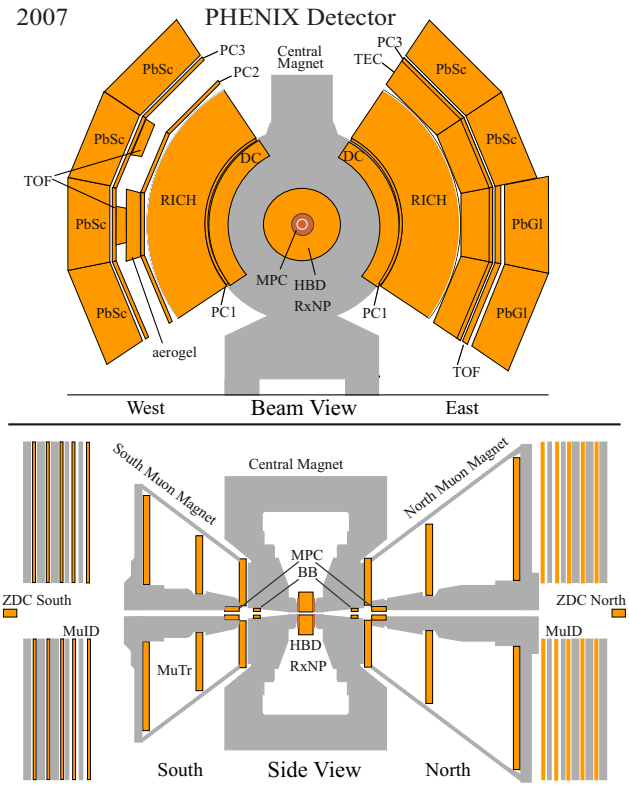}
\caption{(color online) 
The PHENIX detector configuration for RHIC 2007 data-taking 
period.  The beam-beam counters (BBC) are labeled as BB.
} 
\label{Fig:PHENIX}
\end{figure}

Figure~\ref{Fig:PHENIX} shows a schematic diagram of the PHENIX 
detector.  The upper part is a beam-axis view of the two 
central spectrometer arms (West and East), covering the 
pseudorapidity region of $|\eta|<$ 0.35.  Below that is a side 
view showing the two forward-rapidity muon arms (South and 
North) and the global detectors.  A detailed description of the 
complete set of detectors can be found 
elsewhere~\cite{Adcox:2003zm}.

The physics analysis presented here
employed the tracking system [drift chamber (DC) and three layers of
multiwire proportional chambers (PC1, PC2, and PC3)], the West arm
time-of-flight detector (TOFw), the aerogel \v{C}erenkov counter (ACC), the
ring imaging \v{C}erenkov counter (RICH), the beam-beam counters (BBC), the
reaction-plane detector (RxNP), and the muon piston calorimeter
(MPC).  Below, we give a brief description of each of these detector
sub-systems and their role in the present analysis.

\subsection{Global Detectors}
\label{s:global} 

The BBC are located at
$\pm$144~cm from the nominal interaction point along the beam line
in the pseudorapidity region 3.0~$<|\eta|<$~3.9. Each BBC
comprises 64 \v{C}erenkov telescopes, arranged radially around the beam
line. The BBCs provide the main collision trigger for the
experiment and are used in the determination of the collision
vertex position along the beam axis ($z$-vertex) and the
centrality of the collisions. They also provide the start time for
the time-of-flight measurement with timing resolution of
$\sigma_{\rm BBC} = 37$ ps.

The RxNP~\cite{RxNP:2010cc} was
installed in PHENIX before the 2007 data-taking period. It
is located at $\pm$38~cm from the nominal interaction point and
has full azimuthal coverage. Each RxNP comprises two rings of
plastic scintillator paddles, with each paddle subtending $\Delta
\phi = \pi/6$. The inner and outer segments cover the
pseudorapidity ranges 1.0~$<|\eta|<$~1.5 and 1.5~$<|\eta|<$~2.8,
respectively. The RxNP is the main detector used in the
event-plane determination for this analysis. The event-plane
resolution (${\rm Res}(\Psi)$)~\cite{Poskanzer:1998yz}, which is
used as a correction to the $v_{2}$ measurement, is defined as
\begin{equation}
{\rm Res}(\Psi)=\left\langle\cos(2(\Psi-\Psi_{RP})\right\rangle.
\end{equation}
Here the bracket $\langle \rangle$ indicates an average over all events,
$\Psi_{RP}$ is true reaction plane (which is defined by
the directions of interaction parameter and beam),
and $\Psi$ is
the event plane (which is measured by the detector event by event).
A larger value of ${\rm Res}(\Psi)$ corresponds to a better
measurement of the event-plane. In a given event, the event-plane
resolution depends on the charged-particle multiplicity and the
size of the azimuthal anisotropy signal; thus the resolution is
centrality dependent. A resolution of up to 73\% is achieved for
midcentral events.

The MPC are electromagnetic
calorimeters situated at $\pm$223~cm from the nominal interaction
point inside the cylindrical openings at the front of the muon
magnet pistons~\cite{Chiu:2007zy} and have 2$\pi$ azimuthal
acceptance. The pseudorapidity coverage is about 3.0~$<\eta<$~3.8
for the north side and -3.7~$<\eta<$~-3.1 for the south side. The
MPCs are comprised of 220 modules in the north piston hole and 192
in the south with PbWO$_4$ crystals and Avalanche Photodiode
readouts, and can detect both charged and neutral particles. In
this analysis, the MPCs were used for event-plane determination.
Although the event-plane resolution (up to 50\% in midcentral
collisions) is lower than that achieved with the RxNP, the MPCs
provide an important systematic check on the RxNP measurement due
to their larger pseudorapidity separation from the central
spectrometer and therefore smaller nonflow effects on the $v_2$
measurement.

\subsection{Tracking and Particle Identification Detectors}

\label{s:detectors} The charged particle momentum is reconstructed
in the tracking system comprised of the DC located
outside of an axially-symmetric magnetic field at a radial
distance between 2.0~m and 2.4~m followed by the PC1 
with pixel-pad readout.  The pattern
recognition in the DC is based on a combinatorial Hough transform
in the track bend plane. A track model based on a field-integral
look-up table determines the charged particle momentum, the path
length to the TOFw and a projection of the
track to the outer detectors. The momentum resolution in this data
set was estimated to be $\delta p/p \approx 1.3\% \oplus
1.2\%\times p$ (GeV/$c$), where the first term represents multiple
scattering up to the DC and the second term is due to
the DC spatial resolution.  The momentum resolution is worse
than that of previous data set is due to the weaker magnet
configuration.

The tracks are matched to hits registered in the second and third
layers of the pad chambers, PC2 and PC3, which are located at radial
distances of 4.19~m and 4.98~m from the interaction point. Thus, the
contribution of tracks originating from decays and
$\gamma$-conversions is reduced.

To improve the track purity further, we employ the RICH, which 
is a threshold gas \v{C}erenkov detector
located in the radial region 2.5~m $<r<$ 4.1~m. The \v{C}erenkov radiator
gas (CO$_{2}$) at atmospheric pressure has an index of refraction
$n =$ 1.000410 ($\gamma_{th} = 35$), which corresponds to a
momentum threshold of 20~MeV/$c$ for an electron and 4.65~GeV/$c$
for a pion. The RICH provides a veto for the electrons and
positrons, which are predominantly pairs resulting from
$\gamma$-conversions and Dalitz decays.

\begin{figure}[htbp]
\includegraphics[width=0.97\linewidth]{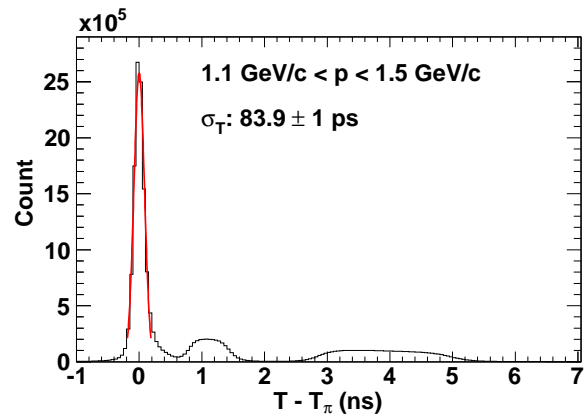}
\caption{(color online) 
Timing difference $T-T_{\pi}$, the difference between the
measured time in the TOFw and the time calculated assuming each
candidate track is a pion.}\label{f:ttof}
\end{figure}

The primary PID used in this analysis
is the TOFw, which is located at a radial distance of
4.81~m from the interaction point and covers the pseudorapidity
range $|\eta|<$ 0.35 and $\delta \phi =$ 22$^\circ$ in azimuth.
The TOFw was built using multigap resistive plate chamber
technology (MRPC)~\cite{Mrpc:2003aa} and installed in PHENIX
before the 2007 data taking period. The MRPCs have six gas gaps
formed by layered glass plates with thickness of 550~$\mu$m,
separated by 230~$\mu$m-thick monofilament fishing line. The MRPCs
are positioned in a gas volume and operated with a gas mixture of
95\% R134a and 5\% isobutane (C$_{4}$H$_{10}$), and bias voltage
of 14 kV. The TOFw system is composed of 128 MRPC modules each of
which has four signal strips of size 37 $\times$ 2.8~cm$^{2}$ and
separation of 0.3~cm. The readout~\cite{Llope:2008eh} is
double-sided, which allows for hit positioning along the direction
of the strip to be determined using the timing difference between
the signals with resolution of the order 1~cm. The other two hit
coordinates are determined using the global position of the strips
within PHENIX. The average time measured on both sides of the
strips provides the stop time for the time-of-flight measurement.
The timing resolution of the BBC-TOFw system was determined by
selecting charged tracks (see Section~\ref{s:events_and_tracks})
with momentum in the range 1.1~GeV/$c$~$<p<$~1.5~GeV/$c$ and
examining the timing difference between the measured flight-time
and the time which is expected under the assumption that the
particles are pions. The resulting time distribution is shown in
Fig.~\ref{f:ttof}. Since the pions dominate the total yield in
this momentum region, a narrow peak centered around
$t-t_{{\rm expected}} \approx 0$ is observed. The other two
broad peaks in Fig.~\ref{f:ttof} correspond to kaons and
protons. A Gaussian distribution is fit to the pion peak and
yields a resolution of $\sigma_{BBC-TOFw}= 84\pm1$~ps for the
BBC-TOFw system.

The excellent timing resolution allows for 4$\sigma $ separation
in mass-squared reaching up to $p_T = $ 2.5~GeV/$c$ for $\pi/K$, and
up to $p_T=$ 4~GeV/$c$ for $K/p$. The PID is further extended in
$p_T$ by use of asymmetric cuts around the centroids of the
mass-squared distributions.

\begin{table}[htbp]
   \caption{\label{t:dPID} The main characteristic parameters of TOFw and ACC}
    \begin{ruledtabular}\begin{tabular}{ccc}
       & TOFw & ACC \\
      \hline

      $\Delta\eta$ & (-0.35, 0.35)& (-0.35, 0.35)\\
      $\Delta\phi$ (rad) & (-0.061, 0.110)  & (-0.108, 0.156)\\
      &(0.503, 0.674)&\\
      Radial distance (cm)    & 481.36 & 449.4 \\
      number of cells & 512 & 160\\
      cell size (cm$^2$) & 37 x 2.8 & 11.95 x 23.10\\
    \end{tabular}\end{ruledtabular}
 \end{table}

The ACC is used in conjunction with the
TOFw to aid the PID at high $p_T$. It is situated in the West
spectrometer arm in front of the TOFw detector. The ACC is a \v{C}erenkov
radiation detector with a relatively high index of refraction
($n=$ 1.0113, $\gamma_{th}=$ 8.5), which means that light is
produced at relatively low momenta. The threshold for radiation is
1.0~GeV/$c$ for pions, 3.0~GeV/$c$ for kaons, and 6.0~GeV/$c$ for
protons. The combined ACC-TOFw information allows for $\pi/K$
separation up to $p_T=$~4~GeV/$c$, and $K/p$ separation up to
$p_T=$ 6~GeV/$c$. The main characteristic parameters of TOFw and
ACC can be found in the Table~\ref{t:dPID}.

\section{Analysis Method}
\label{s:analysis}
\subsection{Event and Track Selection}
\label{s:events_and_tracks} The results reported here are obtained
from an analysis of 4.8 $\times$ 10$^9$ minimum bias events
obtained during the 2007 running period. The minimum-bias trigger
is defined by a coincidence between North and South BBC signals
and an energy threshold of one neutron in both the North and South
zero-degree calorimeters~\cite{Adcox:2003zm}. The collision vertex
$z$ is constrained to $|z|<$ 30~cm of the origin of the coordinate
system.

Charged tracks are selected based on the track quality information
from the tracking system (DC-PC1). The tracks are then projected
to the outer detectors and confirmed by requiring that the closest
hit to the track projection is within certain spatial windows in
$\phi$ and $z$. The distributions for the distance between the
closest hit and projection in the azimuthal and $z$ directions are
fitted with a double Gaussian function, one Gaussian function is
for the signal distribution and the other for the background. For
$p_T<$~3~GeV/$c$, hits are required to match the TOFw and the PC3
to within 2$\sigma$ from the signal's Gaussian distribution in
$\phi$ and $z$. For $p_T \ge$~3~GeV/$c$, hits are required to
match the PC2 and the PC3 to within 3$\sigma$ and the TOFw to
within 2$\sigma$ in $\phi$ and $z$. Background from
$\gamma$-conversions is further reduced by applying a RICH veto.
For the pions, this veto only works for
$p_T<$~5~GeV/$c$ since pions with $p_{T}$ higher than that will fire the RICH.
To evaluate the residual background, remaining after these
selections, the background-to-signal (B/S) ratios from the double
Gaussian function fitting within the samples selected for the
analysis are examined. For $p_T<$~3~GeV/$c$ the background
comprises less than 1\% of the selected tracks. At higher $p_T$
the background increases, reaching B/S $\approx$ 7\% for
5.5~GeV/$c$~$<p_T<$~6.0~GeV/$c$ in the 0--20\% centrality bin.

\subsection{Particle Identification}
\label{s:PID} 
\begin{figure}[htbp]
\includegraphics[width=0.97\linewidth]{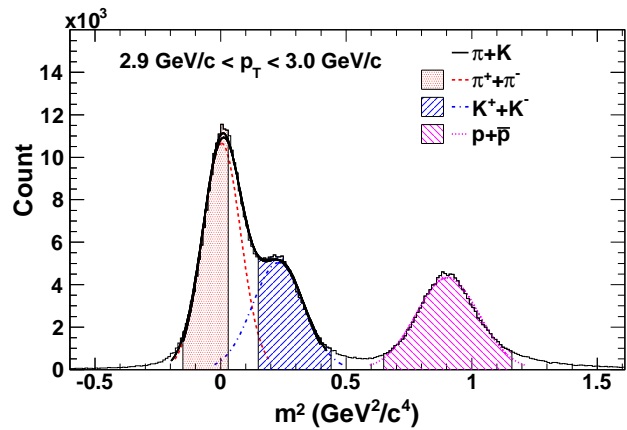}
\caption{(color online) 
The mass-squared distribution measured by TOFw in the
$p_T$ region 2.9~GeV/$c$~$<p_T<$~3.0~GeV/$c$. The hatched areas
show the pion, kaon, and proton selections,
from left to right. The dashed lines show Gaussian fits to the
individual $m^2$ peaks, while the solid line represents a 
combined fit to the $m^2$ distribution
including the pions and kaons.} 
\label{f:tofw3GeV}
\end{figure}

The particles are identified by their mass, based on
measurements of the momentum, the time-of-flight to the TOFw
detector, and the path-length along the trajectory. PID selections
are performed by applying momentum-dependent cuts in mass-squared.
The mass-squared distributions are fit with a 3-Gaussian function
corresponding to pions, kaons, and protons.  The corresponding
widths and centroids are extracted from the data as a function of
transverse momentum. In the calibration process, we ensure that
the centroids of these distributions do not move as a function of
$p_T$ and that the widths vary as expected from the known momentum
and timing resolution of the detector. We then select a sample
from each particle species aiming for at least 90\% purity in PID.
The high purity of the sample will allow us to measure the $v_2$ of
selected particles accurately and minimize the uncertainty resulting
from PID contamination. At lower transverse momenta
($p_T<$~2.5~GeV/$c$), the 2$\sigma$ bands centered around each
particle's $m^2$ do not overlap, thus symmetric cuts,
$m^2_0-2\sigma < m^2 < m^2_0 +2\sigma$, allow for PID with high
purity. In the range 2.5~GeV/$c$~$<p_T<$~3~GeV/$c$, the $\pi/K$
separation is achieved by excluding the particles that lie within
2$\sigma$ of the centroid of the mass-squared distribution of
another particle. This procedure is demonstrated in
Fig.~\ref{f:tofw3GeV}, where the PID selections for $\pi$, $K$,
and $p$ are shown with the hatched areas in the plot. The Gaussian
fits to the individual $m^2$ peaks (dashed-line curves) and the
combined fit to the entire $m^2$ distribution (solid line) are
also shown.

\begin{figure*}[htbp]
\includegraphics[width=0.8\linewidth]{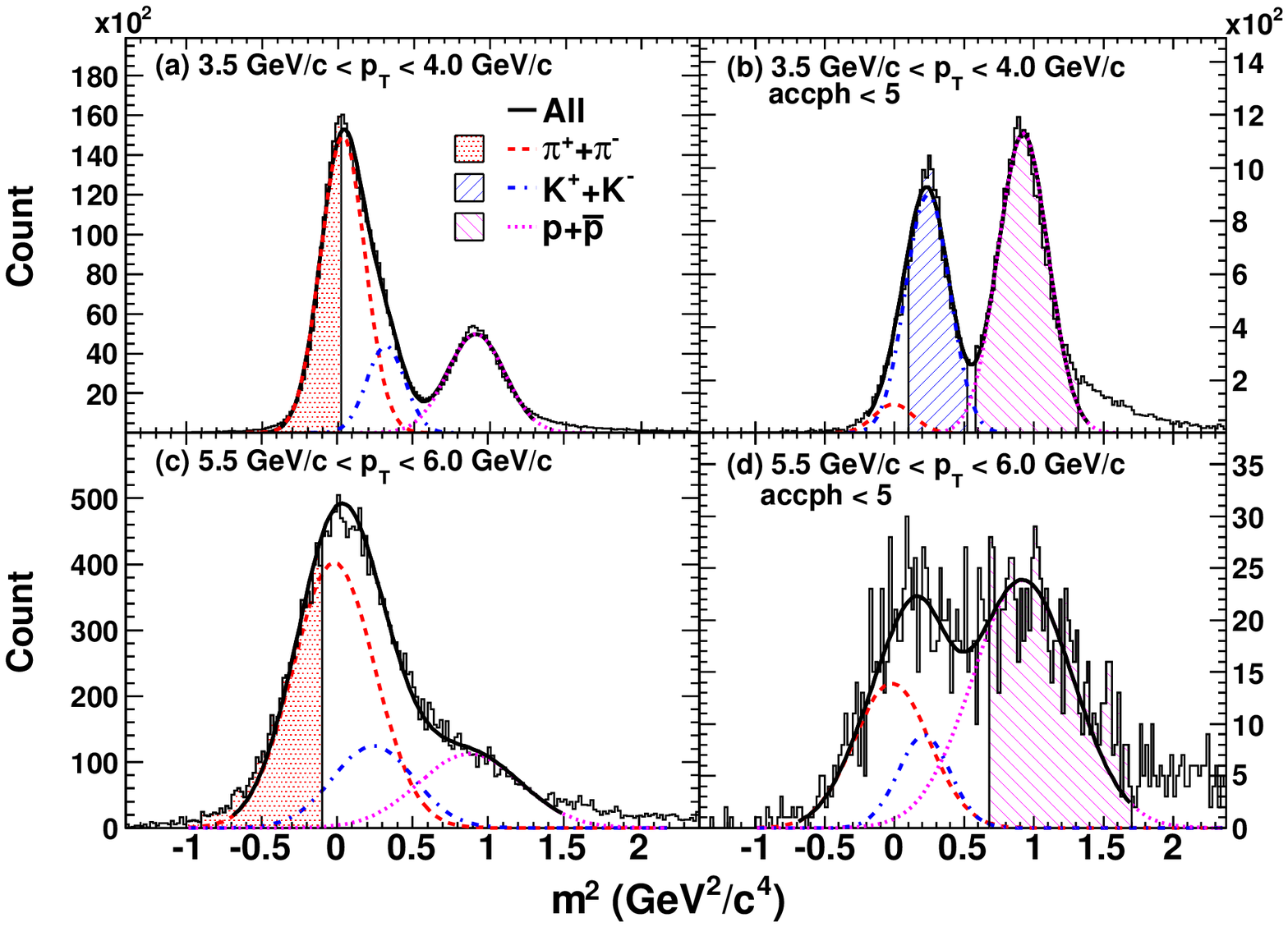}
\caption{(color online) 
The mass-squared distribution in the TOFw without (left
panels) and with (right panels) the ACC
photon yield (accph) cuts for different $p_T$ regions. The
hatched areas show the $m^2$ cuts used for pion, kaon, and proton
selections. The distribution is fit with a 3-Gaussian function
(solid line). The individual Gaussian distributions corresponding
to $\pi$, $K$, and $p$ and are as dashed lines.} \label{f:tofwacc}


\includegraphics[width=0.8\linewidth]{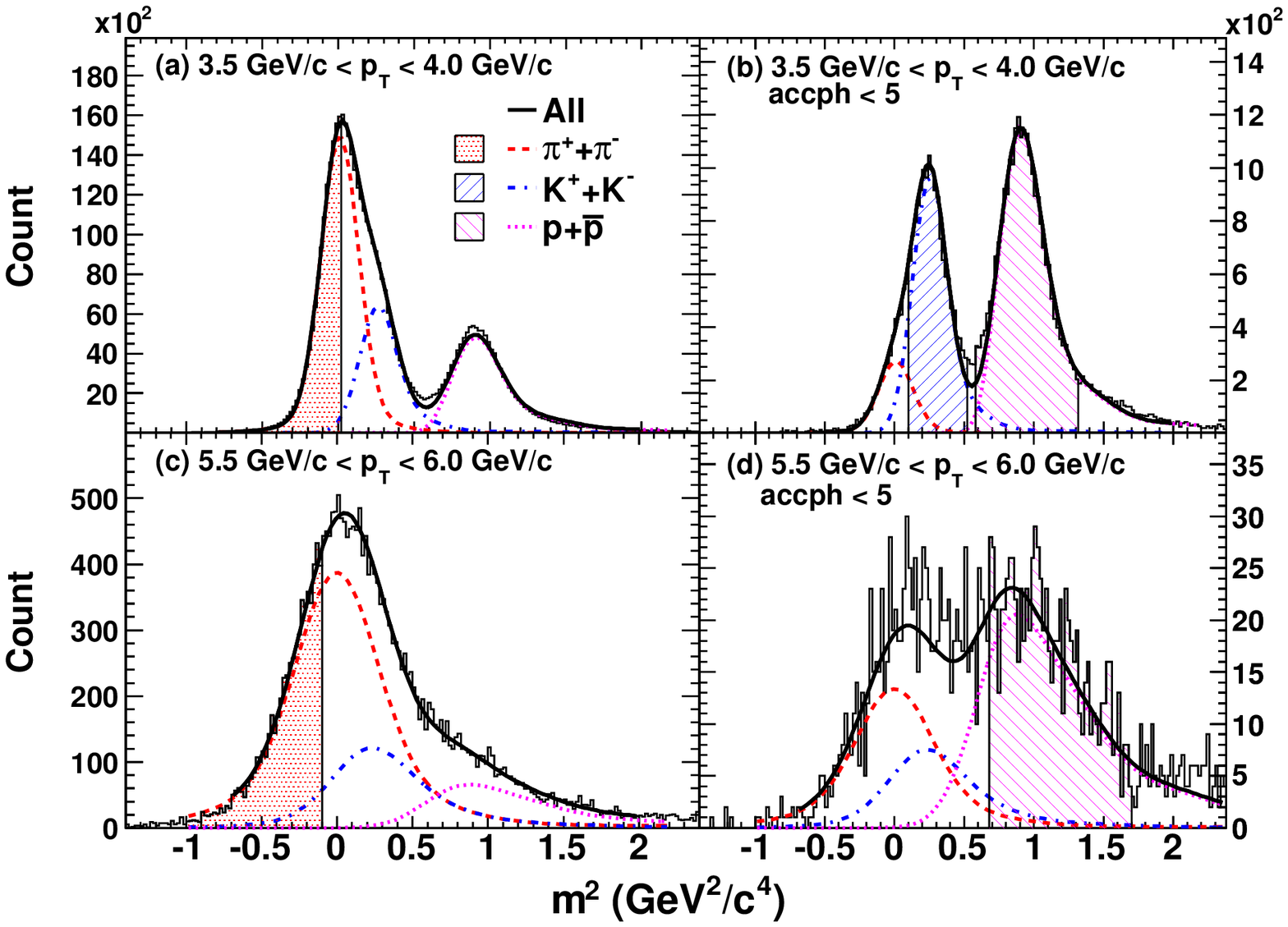}
\caption{(color online) 
The mass-squared distribution in the TOFw without (left
panels) and with (right panels) the ACC 
photon yield (accph) cuts for different $p_T$ regions. The
distribution is fit with the sum of three empirical $m^{2}$
distribution functions that are propagated from sampling a Landau
shape momentum distribution as described in the text (solid line).
The individual distributions corresponding to $\pi$, $K$, and $p$
are shown with dashed lines.} \label{f:landau}
\end{figure*}

At higher transverse momentum 3~GeV/$c$~$<p_T<$~6~GeV/$c$, the lower $m^2$
range of the pion distribution remains relatively unaffected by
contamination from kaons and protons. Therefore, a pion sample with
purity better than 90\% can be selected based on information from
the TOFw alone, by applying the $m^2$ cuts indicated in panels (a) and
(c) of Fig.~\ref{f:tofwacc}, and listed in Table~\ref{t:PID}.

For kaon and proton identification at $p_T>$~3~GeV/$c$, the ACC is
used in conjunction with the TOFw detector, as shown in
Fig.~\ref{f:tofwacc}, panels (b) and (d). The turn-on momenta of
the ACC for pions, kaons, and protons are 1.0~GeV/$c$,
3.0~GeV/$c$, and 6.0~GeV/$c$, respectively. This turn-on is
gradual, with the number of photons registered per photomultiplier
tube (PMT) growing up to 15 for pions, and 10 (kaons and protons)
as the hadrons exceed their respective threshold momentum by
$\approx$~1~GeV/$c$. With this information, the photon yield from
the ACC can be used as a rejection veto based on whether it is
``on'' (accph~$\geq$~5) or ``off'' (accph~$<$~5).
Due to the occupancy effects in the ACC
as well as the spatial resolution of track projection to the ACC, the
pions cannot be rejected completely.
The effect of this veto cut
is demonstrated in Fig.~\ref{f:tofwacc}, panels (b) and (d).
The pion rejection by the ACC in combination with asymmetric $m^2$
cuts, which are indicated here and listed in Table~\ref{t:PID},
allow for kaon and proton PID up to $p_T$ of 4 and 6 GeV/$c$,
respectively.

We use the Gaussian fits to the mass-squared distributions to
estimate the PID purity in the selected $m^{2}$ regions. This is
straightforward at $p_T<$~3~GeV/$c$, where the peaks associated
with each particle are well defined. At higher $p_T$ the
uncertainties are larger, since the pion and kaon peaks merge and
the individual yields are not well constrained. We have checked
that for $p_T>$~4~GeV/$c$, after efficiency corrections, the
$K/\pi$ ratio obtained from our fits is consistent with the
measurements of the $K/\pi$ ($K^{0}_{S}/\pi$) ratio by the STAR
experiment within the statistic and systematic uncertainties. At
$p_{T}$~=~5.22~GeV/$c$, the $K/\pi$ ratio is reported as
0.326~$\pm$~0.013(stat)~$\pm$~0.134(syst) and the $K_S^0/\pi$
ratio is reported as 0.435~$\pm$~0.022(stat)~$\pm$~0.072(syst)
in $p$+$p$ collisions at $\sqrt{s_{NN}}=$~200~GeV; the ratios in
$p$+$p$ and Au+Au collisions are similar~\cite{Ks:2004aa,
Ks:2006bb, Ks:2011cc}. In our study, the kaon contamination in the
pion sample is relatively insensitive to the kaon yield. For
example, if we artificially increase the kaon yield by 30\%, the
contamination in the pion sample increases from 7\% to 9\%.

\begin{table*}[htbp]
   \caption{\label{t:PID}The particle identification cuts in TOFw and
     ACC with PID purity in Au+Au collisions for the centralities 0--20\% and 20--60\%.}
    \begin{ruledtabular}\begin{tabular}{ccccccc}
      Particle & $p_T$ range & TOFw Cut & ACC Cut & \multicolumn{3}{c}{Purity} \\
      & (GeV/$c$) & (GeV/$c^{2}$)$^{2}$ & & & 0--20\% & 20--60\% \\
      \hline

      pion & $<$ 3 & $m^{2}_{\pi} \pm 2 \sigma_{m^{2}_{\pi}}$ & None && $99\%$ & $99\%$\\
      & & veto on $m^{2}_{K} \pm 2 \sigma_{m^{2}_{K}}$ & & & &\\
      & $[3.0,5.0)$ & $[-1.0,0.0]$ & None && $95\%$ & $96\%$\\
      & $[5.0,6.0)$ & $[-1.0,-0.1]$ & None && $91\%$ & $92\%$\\

      kaon & $<$ 3 &  $m^{2}_{K} \pm 2 \sigma_{m^{2}_{K}}$ & None && $98\%$ & $99\%$\\
      & & veto on $m^{2}_{\pi} \pm 2 \sigma_{m^{2}_{\pi}}$ & & & &\\
      & $[3.0 , 3.5)$ & $[0.1,0.5]$ & accph $<$ 5.0 && $94\%$ & $95\%$\\
      & $[3.5 , 4.0)$ & $[0.1,0.5]$ & accph $<$ 5.0 && $91\%$ & $92\%$\\

      proton & $<$ 3 &  $m^{2}_{p} \pm 2 \sigma_{m^{2}_{p}}$ & None && $99\%$ & $99\%$\\
      & $[3.0,4.0)$& $[0.6,1.3]$ & accph $<$ 5.0 && $97\%$ & $98\%$\\
      & $[4.0,5.0)$& $[0.7,1.3]$ & accph $<$ 5.0 && $95\%$ & $96\%$\\
      & $[5.0,6.0)$& $[0.7,1.7]$ & accph $<$ 5.0 && $91\%$ & $92\%$\\
    \end{tabular}\end{ruledtabular}
 \end{table*}
The $m^{2}$ distributions are not strictly Gaussian shape, but
have tails extending to the higher mass region. This effect is not
noticeable at low $p_T$ but comes into prominence at intermediate
and high $p_T$. Hadrons coming from resonance decays may survive
the tracking cuts but will have misreconstructed momentum and
contribute to this high mass tail. The total momentum distribution
of hadrons, including those from resonance decays, is much closer
to a Landau distribution than a Gaussian distribution. To get an
estimate of the possible PID contamination in this case, we have
fit the $m^{2}$ distribution with an empirical function that was
determined by sampling a momentum distribution with a Landau shape
instead of a Gaussian.  This empirical $m^2$ distribution is found
to give a much better fit than a simple 3-Gaussian function and it
gives a good description of the high mass tails. Finally, we
reevaluate the PID contamination with this empirical function. An
example of these fits is shown in Fig.~\ref{f:landau}. The tail
of $m^{2}$ distribution is well described by the empirical pion,
kaon, and proton $m^{2}$ functions which are presented with
different dashed lines. In this case, at high $p_T$ the
contamination from kaons and pions in the proton sample increases
to 9\% from 1\% in the case of the Gaussian fits.

The PID purity for each particle species estimated in different
$p_T$ ranges is listed in Table~\ref{t:PID}. These estimates
reflect the values obtained for the 0--20\% central Au+Au
collisions and are meant to provide lower limits for the
measurements presented here. The purity in more peripheral
collisions was found to be slightly better.

\subsection{Measurement of $v_2$}
\label{s:v2} The measurement of the anisotropy parameter $v_2$
aims to determine the event-by-event particle azimuthal
correlation with the reaction plane of the collision. The true
reaction plane, which is defined as the plane formed by the impact
parameter $b$ and the beam direction, is not known experimentally.
In addition, there exist other sources of correlations in azimuth,
such as the correlations from resonance decays, jets, and quantum
effects. These correlations, which are not related to the reaction
plane, are called nonflow correlations. The goal is to determine
the second coefficient in the Fourier expansion $v_2$ of the
particle azimuthal distribution with respect to the reaction plane
with minimal effects from nonflow correlations. To estimate the
reaction plane angle $\Psi_{RP}$, we employ the event-plane
method~\cite{Poskanzer:1998yz}, in which the second harmonic
azimuthal anisotropy signal determines the event-plane angle
$\Psi$ based on hits registered in one of the event-plane
detectors: RxNP or MPC. For an ideal detector, the measured
distribution of event-plane angles should be isotropic. However,
the actual measurement is usually affected by finite acceptance
and nonuniform efficiencies. We apply a standard event-plane
flattening
technique~\cite{Poskanzer:1998yz,Adler:2003kt,Adler:2005rg,Afanasiev:2007tv}
to remove the residual nonuniformities in the distribution of
event-plane angles. The accuracy with which the event-plane angle
can be determined depends on the strength of the $v_2$ signal and
the multiplicity of the events in each centrality class. It is
maximal for midcentral events, where both of these quantities are
relatively large. The $v_2(p_T)$ measurement is performed by
correlating the particle azimuthal angle $\varphi$ with the second
harmonic event-plane angle $\Psi$, and correcting the observed
signal for the event-plane resolution as follows:

\begin{equation}
v_2 = \frac{\left\langle\cos(2(\varphi -\Psi))\right\rangle}
{{\rm Res}(\Psi)} \label{v2def}
\end{equation}
Here the brackets $\langle \rangle$ indicate an average over all
particles in all events.

Since the true reaction plane angle is not directly measurable,
the resolution correction is estimated using sub-event
techniques~\cite{Poskanzer:1998yz}. There are several different
options in using the sub-event techniques.  The present analysis
uses the two sub-event and the three sub-event methods. These
methods are compared to evaluate the systematic uncertainties
associated with the event-plane resolution corrections.

\begin{figure}[htbp]
\includegraphics[width=0.97\linewidth]{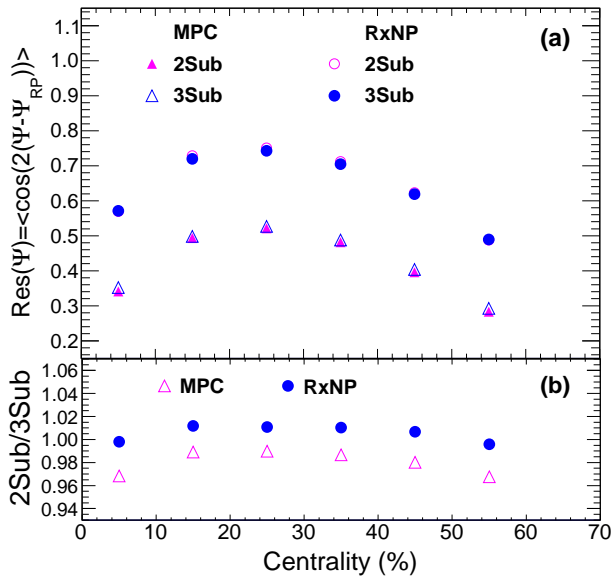}
\caption{(color online) 
Panel (a) shows the event-plane resolution as a function
of centrality for the RxNP and the MPC detectors. Panel (b) shows
the ratio of the event-plane resolution obtained from two
sub-events and three sub-events as a function of
centrality.}\label{f:ev-resolution}
\end{figure}

\begin{figure*}[htbp]
\includegraphics[width=1.0\linewidth]{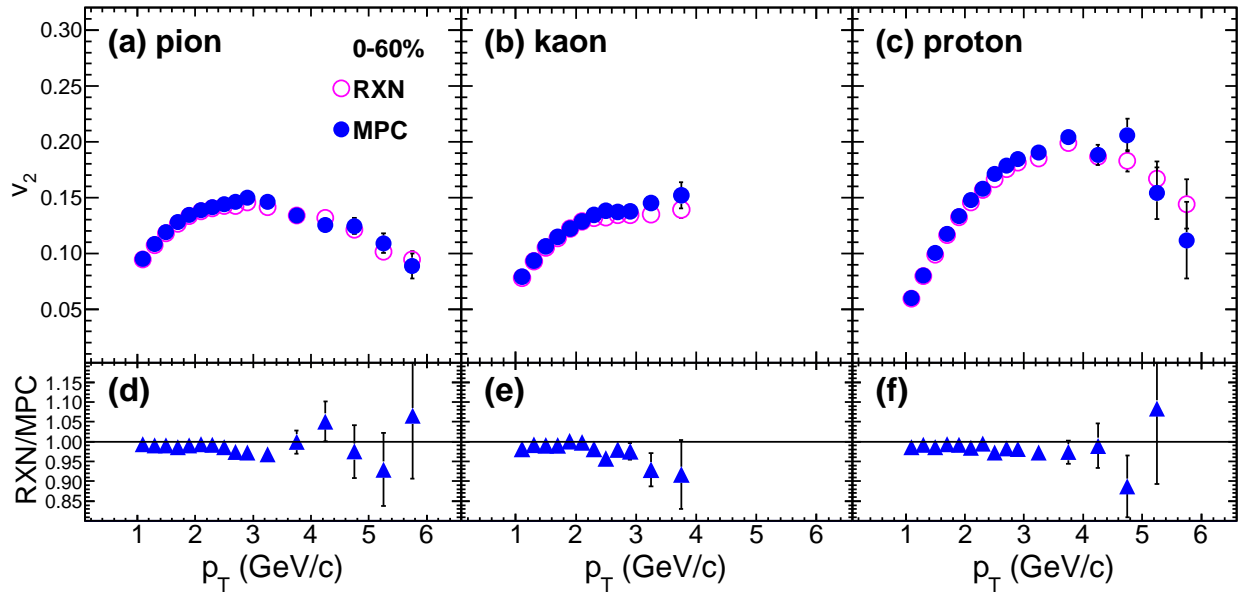}
\caption{(color online) 
The upper panels show the azimuthal anisotropy $v_2(p_T)$
of pions (a), kaons (b), and protons (c) in the 0--60\% centrality
class measured with respect to event-planes determined by the MPC
(closed symbols) or the RxNP (open symbols) detectors. The
event-plane resolution is estimated by the three sub-events
method. The ratio of $v_2$(RxNP) to $v_2$(MPC) is shown in the
lower panels as a function of $p_T$ for pions (d), kaons (e), and
protons (f).}\label{f:3sub-rxn-mpc}
\end{figure*}

The RxNP and the MPC detectors each have two sub-detectors, North
and South, which are positioned symmetrically around the origin of
the nominal collision point with equal acceptance in
pseudorapidity. Thus, they provide a natural two sub-event
division. The correlation between the event-plane angles
determined from the North and South sub-detectors, $\Psi_{N}$ and
$\Psi_{S}$, allows for the estimate of the resolutions corrections
as follows:

\begin{eqnarray}
{\rm Res}(\Psi_{N})={\rm Res}(\Psi_{S})= \sqrt{\langle\cos
2\left(\Psi_{S}-\Psi_{N}\right)\rangle}
  \label{eq:rp_res_2sub}
\end{eqnarray}

\noindent The resolution correction can also be expressed analytically~\cite{Poskanzer:1998yz} as:

\begin{eqnarray}
{\rm Res}(\Psi)=\langle\cos 2\left(\Psi-\Psi_{RP}\right)\rangle =  \nonumber \\
\frac{\sqrt{\pi}}{2}\chi e^{-\frac{\chi^2}{2}}
\left[I_{0}\left(\frac{\chi^2}{2}\right) +
I_{1}\left(\frac{\chi^2}{2}\right)\right], \label{eq:rp_res}
\end{eqnarray}

\noindent where $I_{0}$ and $I_{1}$ are modified Bessel functions.
The parameter $\chi=v_2\sqrt{2M}$, where $M$ is the number of
particles used to determine the event-plane, describes the
dispersion of the flow vector. With the use of
Equation~\ref{eq:rp_res_2sub} and Equation~\ref{eq:rp_res}, we
obtain the sub-event parameters $\chi_{S}$ and $\chi_{N}$.
Subsequently, to optimize the event-plane resolution, the two
sub-events are combined, and the full event parameter is taken as
$\chi=\sqrt{2}\chi_{S}=\sqrt{2}\chi_{N}$. This procedure relies on
the two sub-events being equal in multiplicity and registering the
same size $v_2$ signal, which may not be the case experimentally.
To avoid this uncertainty, we also use a three sub-events
technique to determine the event-plane resolution with
Equation~\ref{eq:rp_res_3sub}~\cite{Poskanzer:1998yz}. To
determine the event-plane resolution of RxNP detector (sub-event
A), we employ information from the North and South portions of the
MPC detector (sub-events B and C). In turn, to estimate the
resolution of the MPC detector, the North and South portions of
the RxNP detector are used to provide sub-events B and C.

\begin{eqnarray}
{\rm Res}(\Psi_{A})=\langle\cos 2\left(\Psi_{A}-\Psi_{RP}\right)\rangle = \nonumber \\
\sqrt{\frac{\langle\cos 2\left(\Psi_{A}-\Psi_{B}\right)\rangle
\langle\cos 2\left(\Psi_{A}-\Psi_{C}\right)\rangle} {\langle\cos
2\left(\Psi_{B}-\Psi_{C}\right)\rangle}}
\label{eq:rp_res_3sub}
\end{eqnarray}

The event-plane resolution for the RxNP (circles) and the MPC
(triangles) detectors obtained with the above procedures are shown
as a function of the event centrality in
Fig.~\ref{f:ev-resolution}(a). The results show the expected
trend, with maximal resolution for the 20--30\% centrality class
where both the event multiplicity and the $v_2$ signal are large,
and a decrease for the more central events (due to lower $v_2$
strength), and more peripheral events (due to smaller
multiplicity).  Figure~\ref{f:ev-resolution}(b) shows the ratio of
the results obtained with the two sub-event and the three
sub-event techniques. The results for the RxNP detector (closed
symbols) agree to within 2\%. A larger difference (up to 4\%) is
observed for the MPC detector (open symbols) which is mainly due
to the asymmetric pseudorapidity coverage of the MPC. The
event-plane resolution from three sub-events method is used to
correct the $v_2$ measurement.

From Fig.~\ref{f:ev-resolution} it is evident that the RxNP
detector has better resolution for the event-plane angle, as well
as smaller systematic uncertainty in the event-plane determination
than the MPC detector. Therefore, it is desirable to use the RxNP
for the $v_2$ measurement. One possible disadvantage of the RxNP
over the MPC detector is the smaller pseudorapidity separation
from the central spectrometer ($|\eta|<$ 0.35), which makes the
$v_2$ measurement more susceptible to nonflow correlations caused
by jets. Since the results presented here aim to study the high
$p_T$ azimuthal anisotropies of identified charged hadrons and the
limits of $n_q$ scaling, it is particularly important to minimize
such effects. To evaluate the nonflow contributions we examine
the $v_2(p_T)$ distributions for pions, kaons, and protons
measured using the MPC and the RxNP detectors independently. For
the $p_{T}<$~6~GeV/$c$, a previous study indicated that nonflow
effects are small for the event-plane measured by the BBC
detectors, which have a pseudorapidity coverage similar to that of
the MPC detectors~\cite{Afanasiev:2009wq}.
Figure~\ref{f:3sub-rxn-mpc} shows the results in the 0--60\%
centrality range for each particle species (upper panels), and the
ratio of the results obtained with the two event-plane detectors
(lower panels). Non-flow correlations are expected to enhance the
measured $v_2$ signal for the detector which is more affected,
especially in the higher $p_T$ range. We do not find any evidence
for a significant increase in nonflow contributions in the
measurement based on the RxNP detector.

Based on these considerations, the results presented in
Section~\ref{s:results} are based on the reaction plane measured
solely by the RxNP, taking advantage of its better event-plane
resolution in comparison to the MPC.

\subsection{Systematic Uncertainties in $v_2$ }
\label{s:sys}
\begin{table*}[htbp]
   \caption{\label{t:sys}Systematic uncertainties given in percent on the $v_2$ measurements.}
    \begin{ruledtabular}\begin{tabular}{cccccccc}
      Error Sources&\hspace{5mm}& Species &\hspace{5mm}& $0-20\%$ && $20-60\%$ & Type \\ \hline
      Event-plane resolution&\hspace{5mm}&&\hspace{5mm}& 2\%&&2\% & C \\
      Event-plane detectors &\hspace{5mm}&&\hspace{5mm}& 3\% in $p_T$ 1--3 GeV/$c$ & \hspace{4mm} & 3\% in $p_T$ 1--5 GeV/$c$& B\\
                       &\hspace{5mm}&&\hspace{5mm}& 5\% in $p_T$ 3--6 GeV/$c$ && 5\% in $p_T$ 5--6 GeV/$c$& \\
      Background &\hspace{5mm}& pion &\hspace{5mm}& 1\% in $p_T$ 1--4 GeV/$c$ && 1\% in $p_T$ 1--4 GeV/$c$& A\\
                       &&&& 4\% in $p_T$ 4--6 GeV/$c$ &  & 3\% in $p_T$ 4--6 GeV/$c$& \\
                 && kaon && 1\% in $p_T$ 1--4 GeV/$c$ && 1\% in $p_T$ 1--4 GeV/$c$& A\\
                 && proton &&1\% in $p_T$ 1--4 GeV/$c$ && 1\% in $p_T$ 1--4 GeV/$c$& A\\
                       &&&&5\% in $p_T$ 4--6 GeV/$c$ && 3\% in $p_T$ 4--6 GeV/$c$& \\
      PID &&pion  &&\multicolumn{3}{c}{negligible in $p_T$ 1--3 GeV/$c$}& A \\
                       &&&&\multicolumn{3}{c}{2\% in $p_T$ 3--6 GeV/$c$}&  \\
          &&kaon  &&\multicolumn{3}{c}{negligible in $p_T$ 1--3GeV/$c$}& A \\
                       &&&&\multicolumn{3}{c}{2\% in $p_T$ 3--4 GeV/$c$}&  \\
          &&proton&&\multicolumn{3}{c}{negligible in $p_T$ 1--3 GeV/$c$}& A \\
                &&&& 3\% in $p_T$ 3--4 GeV/$c$ && 2\% in $p_T$ 3--4 GeV/$c$&  \\
                &&&& 5\% in $p_T$ 4--6 GeV/$c$ && 3\% in $p_T$ 4--6 GeV/$c$&  \\
      Acceptance &&&& 8\% && 3\% & C\\
      and run-by-run &&&&&\\
    \end{tabular}\end{ruledtabular}
 \end{table*}

The systematic uncertainties in the $v_2$ measurement obtained
with the RxNP detector can be broadly characterized according to
the following categories: 1) event-plane resolution corrections;
2) event-plane measured from different detectors; 3) $v_2$ from
background tracks; 4) PID purity; and 5) acceptance and run-by-run
dependencies.

The uncertainties stemming from the event-plane resolution
corrections are independent of particle species and $p_T$. They
are found to be around 2\% for all centralities by studying the
event-plane resolution difference for the RxNP with the two and
three sub-event methods.

The uncertainties from event-planes measured with different
detectors (RxNP, MPC) are found to be independent of the particle
species, by comparing the results from RxNP and MPC. In the
0--20\% centrality class, we assign a 3\% systematic uncertainty
for $p_T<$~3~GeV/$c$ and a 5\% systematic uncertainty for
$p_T>$~3~GeV/$c$. In the 20--60\% centrality class, we assign a
3\% systematic uncertainty for $p_T<$~5~GeV/$c$ and a 5\%
systematic uncertainty for $p_T>$~5~GeV/$c$.

Background tracks that are not removed by the tracking and PID
selections outlined in Sections~\ref{s:events_and_tracks}
and~\ref{s:PID} may influence the measured $v_2$ if they carry a
signal which is different from the particle of interest. The
background tracks may come from decays, $\gamma$-conversions, or
false track reconstruction. The backgrounds are centrality
dependent, and may also have $p_T$ and hadron species dependence.
A sample of background-dominated tracks was selected based on the
normalized distance between the hits registered in the TOFw
detector and the track projections. Specifically, a
4$\sigma$--10$\sigma$ window in the $z$-direction was utilized.
The azimuthal anisotropy of the background was then measured
following the procedure used for the signal. For $p_T
\approx$~3~GeV/$c$, the $v_2$ of the background is similar to that
of the pion, but it decreases at higher $p_T$ down to about $60\%$
of the $v_2$ of the pion (or 30\% of the $v_2$ of the proton) for
$p_T \approx$~6~GeV/$c$ in the 0--20\% centrality class. For
pions, the resulting systematic uncertainties in $v_2$ are of the
order $1\%$ for $p_T<$~4~GeV/$c$ and reach up to $4\%$ ($3\%$) for
$p_T \approx$~6~GeV/$c$ for centrality 0--20\% (20--60\%). For
protons, the resulting systematic uncertainties in $v_2$ are of
the order $1\%$ for $p_T<$~4~GeV/$c$ and reach up to $5\%$ ($3\%$)
for $p_T \approx$~6~GeV/$c$ for centrality 0--20\% (20--60\%).

The systematic uncertainties in $v_2$ resulting from hadron
misidentification are based on the PID purity estimates listed in
Table~\ref{t:PID} and the size of $v_2$ of each species. For
example, at $p_{T}$ = 6.0 GeV/$c$ for 0--20\% centrality, the
protons purity is around 91\% and the $v_2$ of pions and kaons are
around 50\% of that of the proton. We assign a 5\% systematic
uncertainty attributable to this effect. For $p_T<$~3~GeV/$c$, the
uncertainties in $v_2$ due to PID contamination are negligible for
all particle species. At higher $p_T$ the uncertainties in $v_2$
remain below $\approx$ 2\% for kaons and pions; for protons with
$p_T>$~4~GeV/$c$ these uncertainties reach up to
$\approx$~5\%~(3\%) for centrality 0--20\% (20--60\%).

Additional systematic checks were performed using different
subsets of the detector, and data obtained with different magnetic
field configurations. Differences of order 8\% (3\%) were found
for the 0--20\% (20--60\%) centrality, which have weak $p_T$ and
particle species dependence.

Table~\ref{t:sys} lists the summary of all these systematic
uncertainties which are categorized by the types:

(A) point-to-point error uncorrelated between $p_T$ bins,

(B) $p_{T}$ correlated, all points move in the same direction but
not by the same factor,

(C) an overall normalization error in which all points move by the
same factor independent of $p_T$.

\begin{figure*}[htbp]
\includegraphics[width=0.7\linewidth]{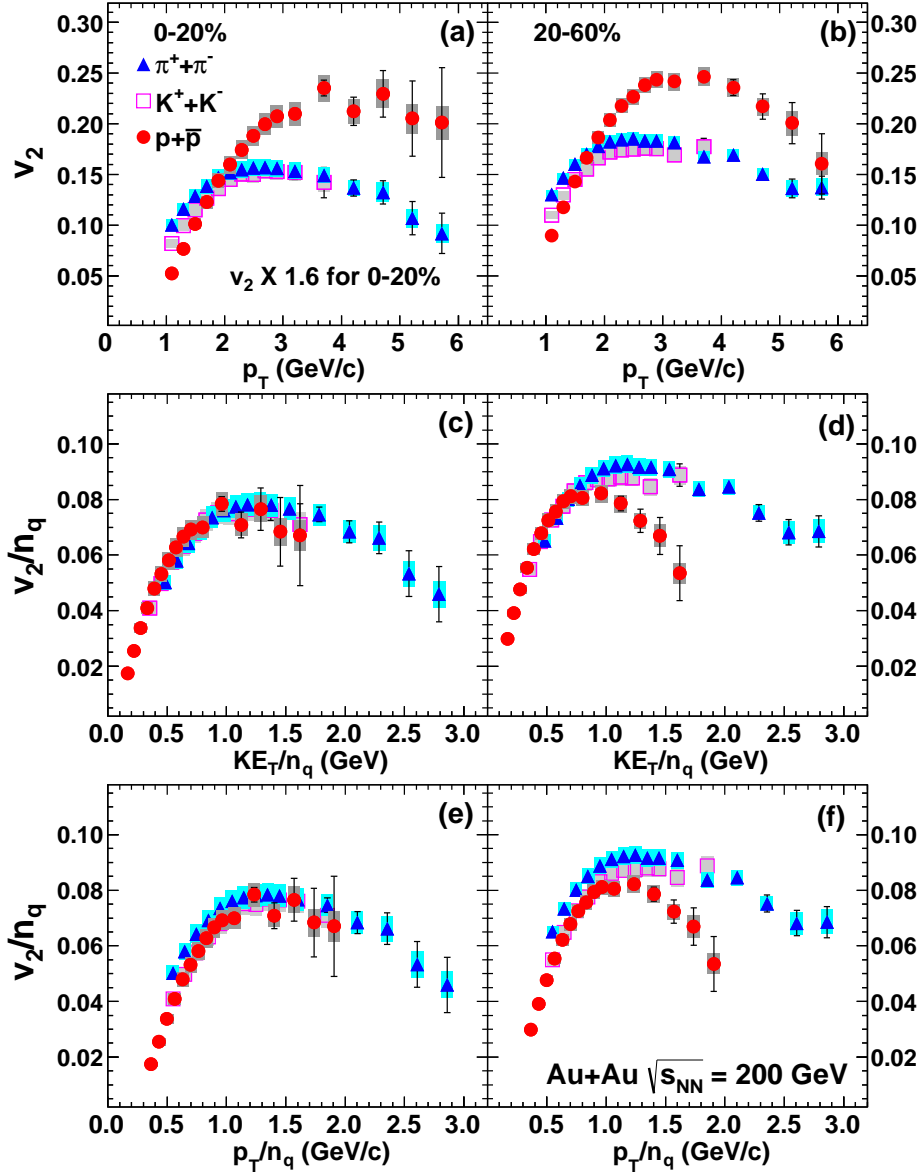}

\caption{(color online) 
Identified hadron $v_2$ in central (0--20\% centrality,
left panels) and midcentral (20--60\%, right panels) Au+Au
collisions at $\sqrt{s_{NN}}=$~200~GeV. Panels (a) and (b) show
$v_2$ as a function of transverse momentum $p_T$. Panels (c) and
(d) show the quark-number-scaled $v_2$ ($v_2/n_q$) as a function
of the kinetic energy per quark, KE$_T/n_q$. Panels (e) and (f)
show $v_2/n_q$ as a function of transverse momentum per quark,
$p_T/n_q$. The $v_2$ of all species for centrality 0--20\% has
been scaled up by a factor of 1.6 for better comparison with
results of 20--60\% centrality. The error bars (shaded boxes)
represent the statistical (systematic) uncertainties. The
systematic uncertainties shown are type A and B only.}\label{f:Fig8}
\end{figure*}

\section{Results and Discussion}
\label{s:results} 

The results for $v_2$ of identified pions, kaons, and protons are 
presented in Fig.~\ref{f:Fig8}; the results in central collisions 
(0--20\%) are presented in panels (a), (c), and (e) and the 
results in noncentral collisions (20--60\%) are presented in 
panels (b), (d), and (f). The symbols representing the different 
particle species are closed triangles for pions, open squares for 
kaons, and closed circles for protons. In order to better compare 
between two centralities, the $v_2$ of all species in the 0--20\% 
centrality has been scaled up by a factor of 1.6. The error bars 
(shaded boxes) represent the statistical (systematic) 
uncertainties. The systematic uncertainties shown are type A and B 
only. Not shown are the type C systematic uncertainties, which are 
from the event-plane resolution, geometrical acceptance, and 
run-by-run dependence are around 8.5\% (3.5\%) for 0--20\% 
(20--60\%) centrality for all species at all values of $p_T$.

Panels (a) and (b) of Fig.~\ref{f:Fig8} show $v_2(p_T)$. For
both centrality selections, the $v_2$ values of pions and kaons
are very similar in intermediate $p_T$ range (2--4 GeV/$c$),
where the measured $v_2$ is maximal and is relatively independent
of transverse momentum. Above $p_T \approx$~4~GeV/$c$ the pion
$v_2$ gradually decreases to a value which is comparable to the
signal measured at $p_T \approx$~1~GeV/$c$. In contrast, the
proton $v_2(p_T)$ has a shape which is centrality dependent. In
central collisions (0--20\%) the proton $v_2$ rises up to $p_T
\approx$~3.5~GeV/$c$ and then saturates at a value higher than the
$v_2$ of pions. For noncentral collisions, the behavior is
different: a decrease is observed in the proton $v_2$ for
$p_T>$~4~GeV/$c$ leading to near equal $v_2$ signals for pions and
protons at $p_T \approx$~6~GeV/$c$.

\begin{figure*}[htbp]
\includegraphics[width=0.7\linewidth]{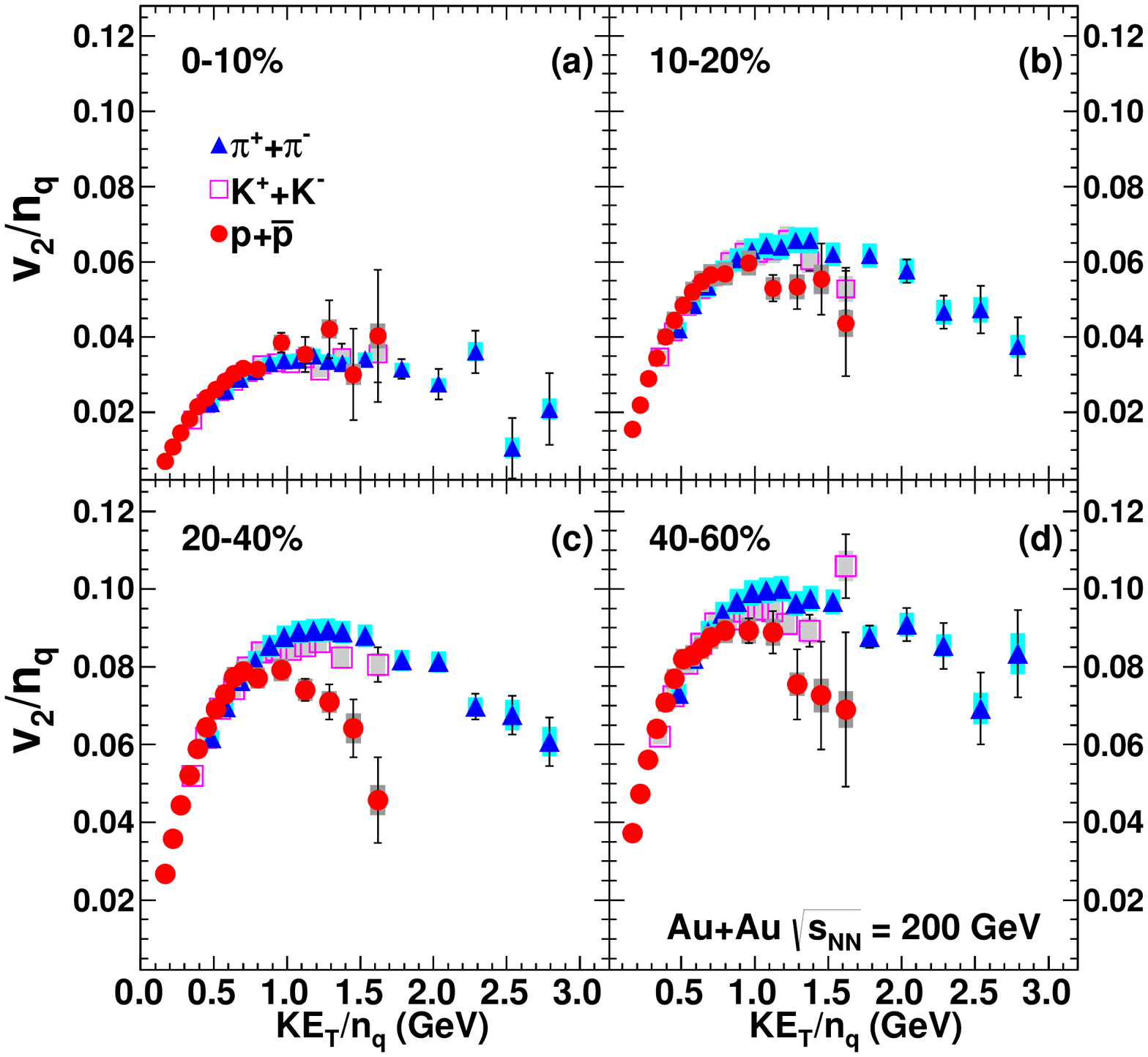}
\caption{(color online) 
The quark-number-scaled $v_2$ ($v_2/n_q$) of identified
hadrons are shown as a function of the kinetic energy per quark,
KE$_T/n_q$ in 0--10\% centrality (panel (a)), 10--20\% (panel
(b)), 20--40\% (panel (c)), and 40--60\% centrality (panel (d)) in
Au+Au collisions at $\sqrt{s_{NN}}=$~200~GeV. The error bars
(shaded boxes) represent the statistical (systematic)
uncertainties. The systematic uncertainties shown are type A and B only.}\label{f:Fig9}

\includegraphics[width=0.7\linewidth]{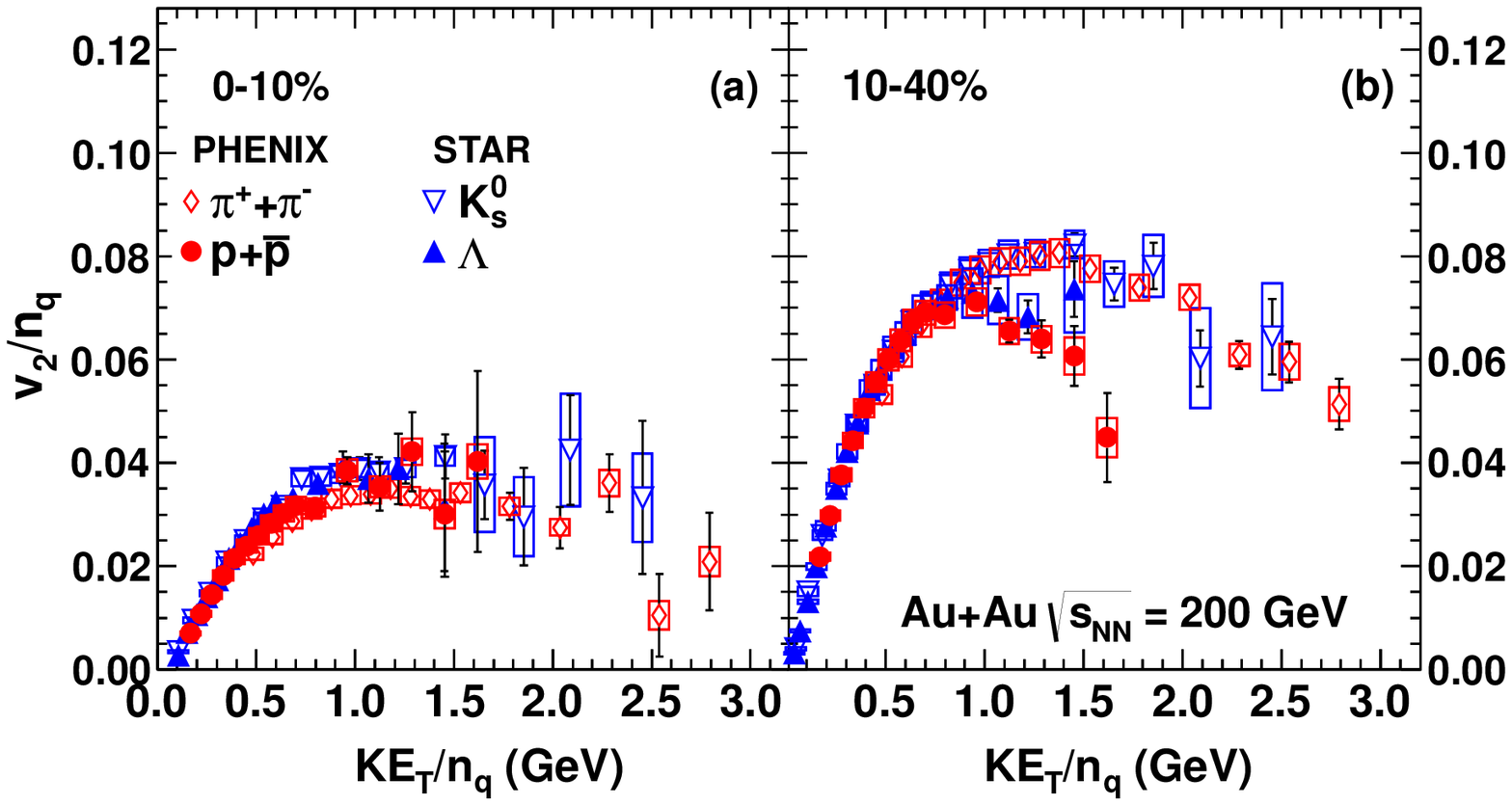}
\caption{(color online) 
The quark-number-scaled $v_2$ ($v_2/n_q$) of identified
hadrons are shown as a function of the kinetic energy per quark,
KE$_T/n_q$ in 0--10\% centrality (panel (a)) and 10--40\%
centrality (panel (b)) in Au+Au collisions at
$\sqrt{s_{NN}}=$~200~GeV. The $v_2$ of $\Lambda$ and $K^{0}_{S}$
are measured by STAR collaboration~\cite{Abelev:2008ed}. The error
bars (open boxes) represent the statistical (systematic)
uncertainties. The systematic uncertainties shown on the results
from this study are type A and B only.}\label{f:Fig10}
\end{figure*}

The use of the KE$_T$ variable was introduced in
Reference~\cite{Adare:2006ti}, which is found to better represent
the number of quark scaling behavior than $p_{T}$ at lower
$p_{T}$. In panels (c) and (d) of Fig.~\ref{f:Fig8} the $v_2$
signals have been scaled by the number of constituent quarks $n_q$
in the hadrons and are shown as a function of the transverse
kinetic energy per quark KE$_T/n_q$. A very different behavior is
observed in central (Fig.~\ref{f:Fig8}(c)) and in noncentral
(Fig.~\ref{f:Fig8}(d)) collisions. In the measured $p_T$ range,
a universal behavior is seen in the central collisions within the
statistical and systematic uncertainties, but not in the
noncentral collisions, where the $v_2/n_q$ of protons falls below
that of the mesons for KE$_T/n_q \ge$~1~GeV. This is the range
where the proton $v_2(p_T)$ begins falling in noncentral
collisions but remains relatively constant in central collisions.

On the other hand, it is widely accepted that the relevant scaling
variable for quark-recombination is the transverse momentum per
quark, since it is the momentum and not the energy that is
additive in the recombination models. Therefore, to examine the
$n_q$ scaling in the recombination regime we show the
quark-number-scaled $v_2$ as a function of $p_T/n_q$ in panels (e)
and (f) of Fig.~\ref{f:Fig8}. For central collisions
(Fig.~\ref{f:Fig8}(e)), the universal behavior appears to
remains valid within the statistical and systematic uncertainties.
Since the changes in $v_2$ are relatively small at higher $p_T$,
shifting the $x$-axis from KE$_T/n_q$ to $p_T/n_q$ does not change
the shape of the curves significantly. For noncentral collisions
(Fig.~\ref{f:Fig8}(f)), the proton data are systematically below
the pion data at all $p_T/n_q$, although they are at the edge of
the systematic uncertainties for $p_T/n_q\le$~1.3~GeV/$c$, which
corresponds to KE$_T/n_q\le$ 1 GeV/$c$. We note that despite this
systematic offset, the $n_q$ scaling makes the shape of the pion
and proton curves very similar below the breaking point. Above
that point, quark recombination is clearly violated.

Some model calculations~\cite{Hwa:0801aa} have shown that the
breaking of $n_q$ scaling occurs at the transition between
purely thermal and thermal+shower recombination.  In the
50--60\% centrality class this can happen for values of KE$_T$
as low as KE$_T/n_{q} \approx$~0.5~GeV, while in the 0--5\% centrality
class this occurs at values as high as KE$_T/n_{q} \approx$~1.5~GeV.
Similar features have been observed in the data presented in this paper.
On the other hand, for pions and protons, the nuclear
modification factors ($R_{\rm AA}$),
which are used to quantify the amount of partonic energy loss in the medium,
have been found to be consistent with each other for
$p_T>$~5~GeV/$c$~\cite{starpid:2006aa,Adare:2011ab,Belmont:2009fm,Ks:2011cc}.
This indicates that a simple interplay between recombination and jet
energy loss is not enough to explain the $v_2$ and $R_{\rm AA}$ of
pions and protons in Au+Au collisions in this $p_T$ region.
Additional considerations may include the nonAbelian nature of
jet energy loss~\cite{Wang:1998ab}, the quark versus gluon
fragmentation production of pions and
protons~\cite{Albino:2005me,Albino:2008fy,Daniel:2007aa}, and
jet chemistry effects such as enhanced parton
splitting~\cite{Sapeta:2008aa} and jet
conversion~\cite{Fries:2009aa}.
Detailed model calculations that take all of these effects into account
are not yet available, and it is an open question whether
doing so is enough for an adequate interpretation of the
$p_T$ $v_2$ and $R_{\rm AA}$ of pions and protons.

To further investigate the centrality dependence of the $n_q$ 
scaling breaking, results with finer centrality bins are shown in 
Fig.~\ref{f:Fig9}. The quark number scaled $v_2$ ($v_2/n_q$) of 
pions, kaons, and protons are shown as a function of the kinetic 
energy per quark KE$_T/n_q$ in 0--10\% (panel (a)), 10--20\% 
(panel (b)), 20--40\% (panel (c)), and 40--60\% centrality (panel 
(d)). The error bars (shaded boxes) represent the statistical 
(systematic) uncertainties. The systematic uncertainties shown are 
type A and B only.  Not shown are the type C systematic 
uncertainties, which are from the event-plane resolution, 
geometrical acceptance, and run-by-run dependence are around 
10.5\% (3.5\%) for 0--10\% (40--60\%). These results with finer 
centrality bins show that the breaking of $n_q$ scaling has a 
clear centrality dependence.

We also compare our results with the existing $v_2$ results for
$K^{0}_{S}$ and $\Lambda$ as measured by the STAR collaboration
using the event-plane method~\cite{Abelev:2008ed} in the 0--10\%
and 10--40\% centrality classes, which are shown in panel (a) and
panel (b) of Fig.~\ref{f:Fig10}, respectively. Since the
event-plane and particles are measured in the same rapidity gap by
the STAR detector in their event-plane method, the $v_2$ values
from STAR measurements are expected to be systematically larger
than those measured by PHENIX~\cite{Adler:2005rg,Abelev:2008ed}
due to nonflow effects.  In the 0--10\% centrality class, the
$v_{2}$ of pions and protons in this study are systematically
lower than the $v_{2}$ of $K^{0}_{S}$ and $\Lambda$ by 17\%
independent of $p_T$, but they are within the systematic
uncertainties. The $n_{q}$ scaling appears to hold in this
centrality class for each particle species. In the 10--40\%
centrality class, the $v_{2}$ of pions and protons are consistent
with that of $K^{0}_{S}$ and $\Lambda$ in the overlapping KE$_T$
region. While the presence of the scaling breaking is not clear in
the $K^{0}_{S}$ and $\Lambda$ results, the improved precision and
extended KE$_{T}$ reach of the present study unambiguously
demonstrates the breaking of $n_q$ scaling in this centrality
class.

\section{Summary}
\label{s:summary} We have presented a high-statistics study of
baryon and meson azimuthal anisotropy $v_2$ measured up to $p_T$
of 6~GeV/$c$ as a function of centrality in
$\sqrt{s_{NN}}=$~200~GeV Au+Au collisions. The $n_q$ scaling is
found to exhibit strong dependence on the collision centrality.
Significant deviations from $n_q$ scaling are found in noncentral
collisions,
starting from the 10--20\% centrality class,
as KE$_{T}/n_{q}>$ 0.7 GeV. These results indicate that parton
fragmentation and the associated energy loss may play an important
role in generating the azimuthal anisotropy of particle emission.
Conversely, in central collisions, such as the 0--10\% centrality
class, the universal $n_q$ scaling appears to hold to
KE$_{T}/n_{q}=$ 1.5 GeV, supporting parton recombination as the
dominant mode of particle production at intermediate transverse
momentum in central Au+Au collisions at top RHIC energy.


%

\section*{ACKNOWLEDGMENTS}   

We thank the staff of the Collider-Accelerator and Physics
Departments at Brookhaven National Laboratory and the staff of
the other PHENIX participating institutions for their vital
contributions.  We acknowledge support from the 
Office of Nuclear Physics in the
Office of Science of the Department of Energy, the
National Science Foundation, Abilene Christian University
Research Council, Research Foundation of SUNY, and Dean of the
College of Arts and Sciences, Vanderbilt University (U.S.A),
Ministry of Education, Culture, Sports, Science, and Technology
and the Japan Society for the Promotion of Science (Japan),
Conselho Nacional de Desenvolvimento Cient\'{\i}fico e
Tecnol{\'o}gico and Funda\c c{\~a}o de Amparo {\`a} Pesquisa do
Estado de S{\~a}o Paulo (Brazil),
Natural Science Foundation of China (P.~R.~China),
Ministry of Education, Youth and Sports (Czech Republic),
Centre National de la Recherche Scientifique, Commissariat
{\`a} l'{\'E}nergie Atomique, and Institut National de Physique
Nucl{\'e}aire et de Physique des Particules (France),
Ministry of Industry, Science and Tekhnologies,
Bundesministerium f\"ur Bildung und Forschung, Deutscher
Akademischer Austausch Dienst, and Alexander von Humboldt Stiftung (Germany),
Hungarian National Science Fund, OTKA (Hungary), 
Department of Atomic Energy and Department of Science and Technology (India), 
Israel Science Foundation (Israel), 
National Research Foundation and WCU program of the 
Ministry Education Science and Technology (Korea),
Ministry of Education and Science, Russian Academy of Sciences,
Federal Agency of Atomic Energy (Russia),
VR and the Wallenberg Foundation (Sweden), 
the U.S. Civilian Research and Development Foundation for the
Independent States of the Former Soviet Union, 
the US-Hungarian Fulbright Foundation for Educational Exchange,
and the US-Israel Binational Science Foundation.


%

\begin{thebibliography}{57}
\expandafter\ifx\csname natexlab\endcsname\relax\def\natexlab#1{#1}\fi
\expandafter\ifx\csname bibnamefont\endcsname\relax
  \def\bibnamefont#1{#1}\fi
\expandafter\ifx\csname bibfnamefont\endcsname\relax
  \def\bibfnamefont#1{#1}\fi
\expandafter\ifx\csname citenamefont\endcsname\relax
  \def\citenamefont#1{#1}\fi
\expandafter\ifx\csname url\endcsname\relax
  \def\url#1{\texttt{#1}}\fi
\expandafter\ifx\csname urlprefix\endcsname\relax\def\urlprefix{URL }\fi
\providecommand{\bibinfo}[2]{#2}
\providecommand{\eprint}[2][]{\url{#2}}

\bibitem[{\citenamefont{Adcox et~al.}(2005)}]{Adcox:2004mh}
\bibinfo{author}{\bibfnamefont{K.}~\bibnamefont{Adcox}} \bibnamefont{et~al.}
  (\bibinfo{collaboration}{PHENIX Collaboration}), \bibinfo{journal}{Nucl.
  Phys. A} \textbf{\bibinfo{volume}{757}}, \bibinfo{pages}{184}
  (\bibinfo{year}{2005}).

\bibitem[{\citenamefont{Adams et~al.}(2005)}]{Adams:2005dq}
\bibinfo{author}{\bibfnamefont{J.}~\bibnamefont{Adams}} \bibnamefont{et~al.}
  (\bibinfo{collaboration}{STAR Collaboration}), \bibinfo{journal}{Nucl. Phys.
  A} \textbf{\bibinfo{volume}{757}}, \bibinfo{pages}{102}
  (\bibinfo{year}{2005}).

\bibitem[{\citenamefont{Back et~al.}(2005)}]{Back:2004je}
\bibinfo{author}{\bibfnamefont{B.~B.} \bibnamefont{Back}} \bibnamefont{et~al.}
  (\bibinfo{collaboration}{PHOBOS Collaboration}), \bibinfo{journal}{Nucl.
  Phys. A} \textbf{\bibinfo{volume}{757}}, \bibinfo{pages}{28}
  (\bibinfo{year}{2005}).

\bibitem[{\citenamefont{Arsene et~al.}(2005)}]{Arsene:2004fa}
\bibinfo{author}{\bibfnamefont{I.}~\bibnamefont{Arsene}} \bibnamefont{et~al.}
  (\bibinfo{collaboration}{BRAHMS Collaboration}), \bibinfo{journal}{Nucl.
  Phys. A} \textbf{\bibinfo{volume}{757}}, \bibinfo{pages}{1}
  (\bibinfo{year}{2005}).

\bibitem[{\citenamefont{Shuryak}(2005)}]{Shuryak:2004cy}
\bibinfo{author}{\bibfnamefont{E.~V.} \bibnamefont{Shuryak}},
  \bibinfo{journal}{Nucl. Phys. A} \textbf{\bibinfo{volume}{750}},
  \bibinfo{pages}{64} (\bibinfo{year}{2005}).

\bibitem[{\citenamefont{Gyulassy and McLerran}(2005)}]{Gyulassy:2004zy}
\bibinfo{author}{\bibfnamefont{M.}~\bibnamefont{Gyulassy}} \bibnamefont{and}
  \bibinfo{author}{\bibfnamefont{L.}~\bibnamefont{McLerran}},
  \bibinfo{journal}{Nucl. Phys. A} \textbf{\bibinfo{volume}{750}},
  \bibinfo{pages}{30} (\bibinfo{year}{2005}).

\bibitem[{\citenamefont{Heinz}()}]{Heinz:2009xj}
\bibinfo{author}{\bibfnamefont{U.~W.} \bibnamefont{Heinz}},
  \bibinfo{note}{arXiv:0901.4355 (2009)}.

\bibitem[{\citenamefont{Romatschke}(2010)}]{Romatschke:2009im}
\bibinfo{author}{\bibfnamefont{P.}~\bibnamefont{Romatschke}},
  \bibinfo{journal}{Int. J. Mod. Phys. E} \textbf{\bibinfo{volume}{19}},
  \bibinfo{pages}{1} (\bibinfo{year}{2010}).

\bibitem[{\citenamefont{Teaney}()}]{Teaney:2009qa}
\bibinfo{author}{\bibfnamefont{D.~A.} \bibnamefont{Teaney}},
  \bibinfo{note}{arXiv:0905.2433 (2009)}.

\bibitem[{\citenamefont{Kovtun et~al.}(2005)\citenamefont{Kovtun, Son, and
  Starinets}}]{Kovtun:2004de}
\bibinfo{author}{\bibfnamefont{P.~K.} \bibnamefont{Kovtun}},
  \bibinfo{author}{\bibfnamefont{D.~T.} \bibnamefont{Son}}, \bibnamefont{and}
  \bibinfo{author}{\bibfnamefont{A.~O.} \bibnamefont{Starinets}},
  \bibinfo{journal}{Phys. Rev. Lett.} \textbf{\bibinfo{volume}{94}},
  \bibinfo{pages}{111601} (\bibinfo{year}{2005}).

\bibitem[{\citenamefont{Adler et~al.}(2007)}]{Just:2007ju}
\bibinfo{author}{\bibfnamefont{S.~S.} \bibnamefont{Adler}} \bibnamefont{et~al.}
  (\bibinfo{collaboration}{PHENIX Collaboration}), \bibinfo{journal}{Phys. Rev.
  C} \textbf{\bibinfo{volume}{76}}, \bibinfo{pages}{034904}
  (\bibinfo{year}{2007}).

\bibitem[{\citenamefont{Afanasiev
  et~al.}(2009{\natexlab{a}})}]{Afanasiev:2009iv}
\bibinfo{author}{\bibfnamefont{S.}~\bibnamefont{Afanasiev}}
  \bibnamefont{et~al.} (\bibinfo{collaboration}{PHENIX Collaboration}),
  \bibinfo{journal}{Phys. Rev. C} \textbf{\bibinfo{volume}{80}},
  \bibinfo{pages}{054907} (\bibinfo{year}{2009}{\natexlab{a}}).

\bibitem[{\citenamefont{Adare et~al.}(2010)}]{Adare:2010sp}
\bibinfo{author}{\bibfnamefont{A.}~\bibnamefont{Adare}} \bibnamefont{et~al.}
  (\bibinfo{collaboration}{PHENIX Collaboration}), \bibinfo{journal}{Phys. Rev.
  Lett.} \textbf{\bibinfo{volume}{105}}, \bibinfo{pages}{142301}
  (\bibinfo{year}{2010}).

\bibitem[{\citenamefont{Bass et~al.}(2009)\citenamefont{Bass, Gale, Majumder,
  Nonaka, Qin et~al.}}]{Bass:2008rv}
\bibinfo{author}{\bibfnamefont{S.~A.} \bibnamefont{Bass}},
  \bibinfo{author}{\bibfnamefont{C.}~\bibnamefont{Gale}},
  \bibinfo{author}{\bibfnamefont{A.}~\bibnamefont{Majumder}},
  \bibinfo{author}{\bibfnamefont{C.}~\bibnamefont{Nonaka}},
  \bibinfo{author}{\bibfnamefont{G.-Y.} \bibnamefont{Qin}},
  \bibnamefont{et~al.}, \bibinfo{journal}{Phys. Rev. C}
  \textbf{\bibinfo{volume}{79}}, \bibinfo{pages}{024901}
  (\bibinfo{year}{2009}).

\bibitem[{\citenamefont{Wang}(2001)}]{Wang:2001xn}
\bibinfo{author}{\bibfnamefont{X.~N.} \bibnamefont{Wang}},
  \bibinfo{journal}{Phys. Rev. C} \textbf{\bibinfo{volume}{63}},
  \bibinfo{pages}{054902} (\bibinfo{year}{2001}).

\bibitem[{\citenamefont{Adler et~al.}(2003{\natexlab{a}})}]{Adler:2003kt}
\bibinfo{author}{\bibfnamefont{S.~S.} \bibnamefont{Adler}} \bibnamefont{et~al.}
  (\bibinfo{collaboration}{PHENIX Collaboration}), \bibinfo{journal}{Phys. Rev.
  Lett.} \textbf{\bibinfo{volume}{91}}, \bibinfo{pages}{182301}
  (\bibinfo{year}{2003}{\natexlab{a}}).

\bibitem[{\citenamefont{Adams et~al.}(2004{\natexlab{a}})}]{Adams:2003am}
\bibinfo{author}{\bibfnamefont{J.}~\bibnamefont{Adams}} \bibnamefont{et~al.}
  (\bibinfo{collaboration}{STAR Collaboration}), \bibinfo{journal}{Phys. Rev.
  Lett.} \textbf{\bibinfo{volume}{92}}, \bibinfo{pages}{052302}
  (\bibinfo{year}{2004}{\natexlab{a}}).

\bibitem[{\citenamefont{Adare et~al.}(2007)}]{Adare:2006ti}
\bibinfo{author}{\bibfnamefont{A.}~\bibnamefont{Adare}} \bibnamefont{et~al.}
  (\bibinfo{collaboration}{PHENIX Collaboration}), \bibinfo{journal}{Phys. Rev.
  Lett.} \textbf{\bibinfo{volume}{98}}, \bibinfo{pages}{162301}
  (\bibinfo{year}{2007}).

\bibitem[{\citenamefont{Afanasiev et~al.}(2007)}]{Afanasiev:2007tv}
\bibinfo{author}{\bibfnamefont{S.}~\bibnamefont{Afanasiev}}
  \bibnamefont{et~al.} (\bibinfo{collaboration}{PHENIX Collaboration}),
  \bibinfo{journal}{Phys. Rev. Lett.} \textbf{\bibinfo{volume}{99}},
  \bibinfo{pages}{052301} (\bibinfo{year}{2007}).

\bibitem[{\citenamefont{Abelev et~al.}(2007)}]{Abelev:2007qg}
\bibinfo{author}{\bibfnamefont{B.}~\bibnamefont{Abelev}} \bibnamefont{et~al.}
  (\bibinfo{collaboration}{the STAR Collaboration}), \bibinfo{journal}{Phys.
  Rev. C} \textbf{\bibinfo{volume}{75}}, \bibinfo{pages}{054906}
  (\bibinfo{year}{2007}).

\bibitem[{\citenamefont{Abelev et~al.}(2008)}]{Abelev:2008ed}
\bibinfo{author}{\bibfnamefont{B.}~\bibnamefont{Abelev}} \bibnamefont{et~al.}
  (\bibinfo{collaboration}{STAR Collaboration}), \bibinfo{journal}{Phys. Rev.
  C} \textbf{\bibinfo{volume}{77}}, \bibinfo{pages}{054901}
  (\bibinfo{year}{2008}).

\bibitem[{\citenamefont{Hwa and Yang}(2003)}]{Hwa:2002tu}
\bibinfo{author}{\bibfnamefont{R.~C.} \bibnamefont{Hwa}} \bibnamefont{and}
  \bibinfo{author}{\bibfnamefont{C.~B.} \bibnamefont{Yang}},
  \bibinfo{journal}{Phys. Rev. C} \textbf{\bibinfo{volume}{67}},
  \bibinfo{pages}{034902} (\bibinfo{year}{2003}).

\bibitem[{\citenamefont{Fries et~al.}(2003{\natexlab{a}})\citenamefont{Fries,
  Muller, Nonaka, and Bass}}]{Fries:2003vb}
\bibinfo{author}{\bibfnamefont{R.~J.} \bibnamefont{Fries}},
  \bibinfo{author}{\bibfnamefont{B.}~\bibnamefont{Muller}},
  \bibinfo{author}{\bibfnamefont{C.}~\bibnamefont{Nonaka}}, \bibnamefont{and}
  \bibinfo{author}{\bibfnamefont{S.~A.} \bibnamefont{Bass}},
  \bibinfo{journal}{Phys. Rev. Lett.} \textbf{\bibinfo{volume}{90}},
  \bibinfo{pages}{202303} (\bibinfo{year}{2003}{\natexlab{a}}).

\bibitem[{\citenamefont{Fries et~al.}(2003{\natexlab{b}})\citenamefont{Fries,
  Muller, Nonaka, and Bass}}]{Fries:2003kq}
\bibinfo{author}{\bibfnamefont{R.~J.} \bibnamefont{Fries}},
  \bibinfo{author}{\bibfnamefont{B.}~\bibnamefont{Muller}},
  \bibinfo{author}{\bibfnamefont{C.}~\bibnamefont{Nonaka}}, \bibnamefont{and}
  \bibinfo{author}{\bibfnamefont{S.~A.} \bibnamefont{Bass}},
  \bibinfo{journal}{Phys. Rev. C} \textbf{\bibinfo{volume}{68}},
  \bibinfo{pages}{044902} (\bibinfo{year}{2003}{\natexlab{b}}).

\bibitem[{\citenamefont{Greco et~al.}(2003)\citenamefont{Greco, Ko, and
  Levai}}]{Greco:2003xt}
\bibinfo{author}{\bibfnamefont{V.}~\bibnamefont{Greco}},
  \bibinfo{author}{\bibfnamefont{C.~M.} \bibnamefont{Ko}}, \bibnamefont{and}
  \bibinfo{author}{\bibfnamefont{P.}~\bibnamefont{Levai}},
  \bibinfo{journal}{Phys. Rev. Lett.} \textbf{\bibinfo{volume}{90}},
  \bibinfo{pages}{202302} (\bibinfo{year}{2003}).

\bibitem[{\citenamefont{Molnar and Voloshin}(2003)}]{Molnar:2003ff}
\bibinfo{author}{\bibfnamefont{D.}~\bibnamefont{Molnar}} \bibnamefont{and}
  \bibinfo{author}{\bibfnamefont{S.~A.} \bibnamefont{Voloshin}},
  \bibinfo{journal}{Phys. Rev. Lett.} \textbf{\bibinfo{volume}{91}},
  \bibinfo{pages}{092301} (\bibinfo{year}{2003}).

\bibitem[{\citenamefont{Adler et~al.}(2003{\natexlab{b}})}]{Adler:2003kg}
\bibinfo{author}{\bibfnamefont{S.~S.} \bibnamefont{Adler}} \bibnamefont{et~al.}
  (\bibinfo{collaboration}{PHENIX Collaboration}), \bibinfo{journal}{Phys. Rev.
  Lett.} \textbf{\bibinfo{volume}{91}}, \bibinfo{pages}{172301}
  (\bibinfo{year}{2003}{\natexlab{b}}).

\bibitem[{\citenamefont{Adler et~al.}(2004)}]{Adler:2003cb}
\bibinfo{author}{\bibfnamefont{S.~S.} \bibnamefont{Adler}} \bibnamefont{et~al.}
  (\bibinfo{collaboration}{PHENIX Collaboration}), \bibinfo{journal}{Phys. Rev.
  C} \textbf{\bibinfo{volume}{69}}, \bibinfo{pages}{034909}
  (\bibinfo{year}{2004}).

\bibitem[{\citenamefont{Muller et~al.}(2005)\citenamefont{Muller, Fries, and
  Bass}}]{Muller:2005pv}
\bibinfo{author}{\bibfnamefont{B.}~\bibnamefont{Muller}},
  \bibinfo{author}{\bibfnamefont{R.~J.} \bibnamefont{Fries}}, \bibnamefont{and}
  \bibinfo{author}{\bibfnamefont{S.~A.} \bibnamefont{Bass}},
  \bibinfo{journal}{Phys. Lett. B} \textbf{\bibinfo{volume}{618}},
  \bibinfo{pages}{77} (\bibinfo{year}{2005}).

\bibitem[{\citenamefont{Chiu et~al.}(2008)\citenamefont{Chiu, Hwa, and
  Yang}}]{Hwa:0801aa}
\bibinfo{author}{\bibfnamefont{C.~B.} \bibnamefont{Chiu}},
  \bibinfo{author}{\bibfnamefont{R.~C.} \bibnamefont{Hwa}}, \bibnamefont{and}
  \bibinfo{author}{\bibfnamefont{C.~B.} \bibnamefont{Yang}},
  \bibinfo{journal}{Phys. Rev. C} \textbf{\bibinfo{volume}{78}},
  \bibinfo{pages}{044903} (\bibinfo{year}{2008}).

\bibitem[{\citenamefont{Ferini et~al.}(2009)\citenamefont{Ferini, Colonna,
  Di~Toro, and Greco}}]{Ferini:2008he}
\bibinfo{author}{\bibfnamefont{G.}~\bibnamefont{Ferini}},
  \bibinfo{author}{\bibfnamefont{M.}~\bibnamefont{Colonna}},
  \bibinfo{author}{\bibfnamefont{M.}~\bibnamefont{Di~Toro}}, \bibnamefont{and}
  \bibinfo{author}{\bibfnamefont{V.}~\bibnamefont{Greco}},
  \bibinfo{journal}{Phys. Lett. B} \textbf{\bibinfo{volume}{670}},
  \bibinfo{pages}{325} (\bibinfo{year}{2009}).

\bibitem[{\citenamefont{Dusling et~al.}(2010)\citenamefont{Dusling, Moore, and
  Teaney}}]{Dusling:2009df}
\bibinfo{author}{\bibfnamefont{K.}~\bibnamefont{Dusling}},
  \bibinfo{author}{\bibfnamefont{G.~D.} \bibnamefont{Moore}}, \bibnamefont{and}
  \bibinfo{author}{\bibfnamefont{D.}~\bibnamefont{Teaney}},
  \bibinfo{journal}{Phys. Rev. C} \textbf{\bibinfo{volume}{81}},
  \bibinfo{pages}{034907} (\bibinfo{year}{2010}).

\bibitem[{\citenamefont{Lacey et~al.}(2010)\citenamefont{Lacey, Taranenko, Wei,
  Ajitanand, Alexander et~al.}}]{Lacey:2010fe}
\bibinfo{author}{\bibfnamefont{R.~A.} \bibnamefont{Lacey}},
  \bibinfo{author}{\bibfnamefont{A.}~\bibnamefont{Taranenko}},
  \bibinfo{author}{\bibfnamefont{R.}~\bibnamefont{Wei}},
  \bibinfo{author}{\bibfnamefont{N.}~\bibnamefont{Ajitanand}},
  \bibinfo{author}{\bibfnamefont{J.}~\bibnamefont{Alexander}},
  \bibnamefont{et~al.}, \bibinfo{journal}{Phys. Rev. C}
  \textbf{\bibinfo{volume}{82}}, \bibinfo{pages}{034910}
  (\bibinfo{year}{2010}).

\bibitem[{\citenamefont{Fochler et~al.}(2009)\citenamefont{Fochler, Xu, and
  Greiner}}]{Fochler:2008ts}
\bibinfo{author}{\bibfnamefont{O.}~\bibnamefont{Fochler}},
  \bibinfo{author}{\bibfnamefont{Z.}~\bibnamefont{Xu}}, \bibnamefont{and}
  \bibinfo{author}{\bibfnamefont{C.}~\bibnamefont{Greiner}},
  \bibinfo{journal}{Phys. Rev. Lett.} \textbf{\bibinfo{volume}{102}},
  \bibinfo{pages}{202301} (\bibinfo{year}{2009}).

\bibitem[{\citenamefont{Lacey et~al.}(2009)\citenamefont{Lacey, Ajitanand,
  Alexander, Gong, Jia et~al.}}]{Lacey:2009ps}
\bibinfo{author}{\bibfnamefont{R.~A.} \bibnamefont{Lacey}},
  \bibinfo{author}{\bibfnamefont{N.}~\bibnamefont{Ajitanand}},
  \bibinfo{author}{\bibfnamefont{J.}~\bibnamefont{Alexander}},
  \bibinfo{author}{\bibfnamefont{X.}~\bibnamefont{Gong}},
  \bibinfo{author}{\bibfnamefont{J.}~\bibnamefont{Jia}}, \bibnamefont{et~al.},
  \bibinfo{journal}{Phys. Rev. C} \textbf{\bibinfo{volume}{80}},
  \bibinfo{pages}{051901} (\bibinfo{year}{2009}).

\bibitem[{\citenamefont{Hirano and Nara}(2004)}]{PhysRevC.69.034908}
\bibinfo{author}{\bibfnamefont{T.}~\bibnamefont{Hirano}} \bibnamefont{and}
  \bibinfo{author}{\bibfnamefont{Y.}~\bibnamefont{Nara}},
  \bibinfo{journal}{Phys. Rev. C} \textbf{\bibinfo{volume}{69}},
  \bibinfo{pages}{034908} (\bibinfo{year}{2004}).

\bibitem[{\citenamefont{Dunlop et~al.}(2011)\citenamefont{Dunlop, Lisa, and
  Sorensen}}]{Dunlop:2011cf}
\bibinfo{author}{\bibfnamefont{J.~C.} \bibnamefont{Dunlop}},
  \bibinfo{author}{\bibfnamefont{M.~A.} \bibnamefont{Lisa}}, \bibnamefont{and}
  \bibinfo{author}{\bibfnamefont{P.}~\bibnamefont{Sorensen}},
  \bibinfo{journal}{Phys. Rev. C} \textbf{\bibinfo{volume}{84}},
  \bibinfo{pages}{044914} (\bibinfo{year}{2011}).

\bibitem[{\citenamefont{Adcox et~al.}(2003)}]{Adcox:2003zm}
\bibinfo{author}{\bibfnamefont{K.}~\bibnamefont{Adcox}} \bibnamefont{et~al.}
  (\bibinfo{collaboration}{PHENIX Collaboration}), \bibinfo{journal}{Nucl.
  Instrum. Meth. A} \textbf{\bibinfo{volume}{499}}, \bibinfo{pages}{469}
  (\bibinfo{year}{2003}).

\bibitem[{\citenamefont{Richardson et~al.}(2011)}]{RxNP:2010cc}
\bibinfo{author}{\bibfnamefont{E.}~\bibnamefont{Richardson}}
  \bibnamefont{et~al.} (\bibinfo{collaboration}{PHENIX Collaboration}),
  \bibinfo{journal}{Nucl. Instrum. Meth. A} \textbf{\bibinfo{volume}{636}},
  \bibinfo{pages}{99} (\bibinfo{year}{2011}).

\bibitem[{\citenamefont{Poskanzer and Voloshin}(1998)}]{Poskanzer:1998yz}
\bibinfo{author}{\bibfnamefont{A.~M.} \bibnamefont{Poskanzer}}
  \bibnamefont{and} \bibinfo{author}{\bibfnamefont{S.~A.}
  \bibnamefont{Voloshin}}, \bibinfo{journal}{Phys. Rev. C}
  \textbf{\bibinfo{volume}{58}}, \bibinfo{pages}{1671} (\bibinfo{year}{1998}).

\bibitem[{\citenamefont{Chiu}(2007)}]{Chiu:2007zy}
\bibinfo{author}{\bibfnamefont{M.}~\bibnamefont{Chiu}}
  (\bibinfo{collaboration}{PHENIX Collaboration}), \bibinfo{journal}{AIP Conf.
  Proc.} \textbf{\bibinfo{volume}{915}}, \bibinfo{pages}{539}
  (\bibinfo{year}{2007}).

\bibitem[{\citenamefont{Bonner et~al.}(2003)}]{Mrpc:2003aa}
\bibinfo{author}{\bibfnamefont{B.}~\bibnamefont{Bonner}} \bibnamefont{et~al.},
  \bibinfo{journal}{Nucl. Instr. Meth. A} \textbf{\bibinfo{volume}{508}},
  \bibinfo{pages}{181} (\bibinfo{year}{2003}).

\bibitem[{\citenamefont{Llope et~al.}(2008)}]{Llope:2008eh}
\bibinfo{author}{\bibfnamefont{W.~J.} \bibnamefont{Llope}}
  \bibnamefont{et~al.}, \bibinfo{journal}{Nucl. Instrum. Meth. A}
  \textbf{\bibinfo{volume}{596}}, \bibinfo{pages}{430} (\bibinfo{year}{2008}).

\bibitem[{\citenamefont{Adams et~al.}(2004{\natexlab{b}})}]{Ks:2004aa}
\bibinfo{author}{\bibfnamefont{J.}~\bibnamefont{Adams}} \bibnamefont{et~al.}
  (\bibinfo{collaboration}{STAR Collaboration}), \bibinfo{journal}{Phys. Rev.
  Lett.} \textbf{\bibinfo{volume}{92}}, \bibinfo{pages}{052302}
  (\bibinfo{year}{2004}{\natexlab{b}}).

\bibitem[{\citenamefont{Adams et~al.}()}]{Ks:2006bb}
\bibinfo{author}{\bibfnamefont{J.}~\bibnamefont{Adams}} \bibnamefont{et~al.},
  \bibinfo{note}{(STAR Collaboration), nucl-ex/0601042 (2006)}.

\bibitem[{\citenamefont{Agakishiev et~al.}()}]{Ks:2011cc}
\bibinfo{author}{\bibfnamefont{G.}~\bibnamefont{Agakishiev}}
  \bibnamefont{et~al.}, \bibinfo{note}{(STAR Collaboration), arXiv:1110.0579
  (2011)}.

\bibitem[{\citenamefont{Adler et~al.}(2006)}]{Adler:2005rg}
\bibinfo{author}{\bibfnamefont{S.~S.} \bibnamefont{Adler}} \bibnamefont{et~al.}
  (\bibinfo{collaboration}{PHENIX Collaboration}), \bibinfo{journal}{Phys. Rev.
  Lett.} \textbf{\bibinfo{volume}{96}}, \bibinfo{pages}{032302}
  (\bibinfo{year}{2006}).

\bibitem[{\citenamefont{Afanasiev
  et~al.}(2009{\natexlab{b}})}]{Afanasiev:2009wq}
\bibinfo{author}{\bibfnamefont{S.}~\bibnamefont{Afanasiev}}
  \bibnamefont{et~al.} (\bibinfo{collaboration}{PHENIX Collaboration}),
  \bibinfo{journal}{Phys. Rev. C} \textbf{\bibinfo{volume}{80}},
  \bibinfo{pages}{024909} (\bibinfo{year}{2009}{\natexlab{b}}).

\bibitem[{\citenamefont{Abelev et~al.}(2006)}]{starpid:2006aa}
\bibinfo{author}{\bibfnamefont{B.~I.} \bibnamefont{Abelev}}
  \bibnamefont{et~al.} (\bibinfo{collaboration}{STAR Collaboration}),
  \bibinfo{journal}{Phys. Rev. Lett.} \textbf{\bibinfo{volume}{97}},
  \bibinfo{pages}{152301} (\bibinfo{year}{2006}).

\bibitem[{\citenamefont{Adare et~al.}(2011)}]{Adare:2011ab}
\bibinfo{author}{\bibfnamefont{A.}~\bibnamefont{Adare}} \bibnamefont{et~al.}
  (\bibinfo{collaboration}{PHENIX Collaboration}), \bibinfo{journal}{Phys. Rev.
  C} \textbf{\bibinfo{volume}{83}}, \bibinfo{pages}{064903}
  (\bibinfo{year}{2011}).

\bibitem[{\citenamefont{Belmont}(2009)}]{Belmont:2009fm}
\bibinfo{author}{\bibfnamefont{R.}~\bibnamefont{Belmont}}
  (\bibinfo{collaboration}{PHENIX Collaboration}), \bibinfo{journal}{Nucl.
  Phys. A} \textbf{\bibinfo{volume}{830}}, \bibinfo{pages}{697c}
  (\bibinfo{year}{2009}).

\bibitem[{\citenamefont{Wang}(1998)}]{Wang:1998ab}
\bibinfo{author}{\bibfnamefont{X.~N.} \bibnamefont{Wang}},
  \bibinfo{journal}{Phys. Rev. C} \textbf{\bibinfo{volume}{58}},
  \bibinfo{pages}{2321} (\bibinfo{year}{1998}).

\bibitem[{\citenamefont{Albino et~al.}(2005)\citenamefont{Albino, Kniehl, and
  Kramer}}]{Albino:2005me}
\bibinfo{author}{\bibfnamefont{S.}~\bibnamefont{Albino}},
  \bibinfo{author}{\bibfnamefont{B.~A.} \bibnamefont{Kniehl}},
  \bibnamefont{and} \bibinfo{author}{\bibfnamefont{G.}~\bibnamefont{Kramer}},
  \bibinfo{journal}{Nucl. Phys. B} \textbf{\bibinfo{volume}{725}},
  \bibinfo{pages}{181} (\bibinfo{year}{2005}).

\bibitem[{\citenamefont{Albino et~al.}(2008)\citenamefont{Albino, Kniehl, and
  Kramer}}]{Albino:2008fy}
\bibinfo{author}{\bibfnamefont{S.}~\bibnamefont{Albino}},
  \bibinfo{author}{\bibfnamefont{B.~A.} \bibnamefont{Kniehl}},
  \bibnamefont{and} \bibinfo{author}{\bibfnamefont{G.}~\bibnamefont{Kramer}},
  \bibinfo{journal}{Nucl. Phys. B} \textbf{\bibinfo{volume}{803}},
  \bibinfo{pages}{42} (\bibinfo{year}{2008}).

\bibitem[{\citenamefont{de~Florian et~al.}(2007)\citenamefont{de~Florian,
  Sassot, and Stratmann}}]{Daniel:2007aa}
\bibinfo{author}{\bibfnamefont{D.}~\bibnamefont{de~Florian}},
  \bibinfo{author}{\bibfnamefont{R.}~\bibnamefont{Sassot}}, \bibnamefont{and}
  \bibinfo{author}{\bibfnamefont{M.}~\bibnamefont{Stratmann}},
  \bibinfo{journal}{Phys. Rev. D} \textbf{\bibinfo{volume}{76}},
  \bibinfo{pages}{074033} (\bibinfo{year}{2007}).

\bibitem[{\citenamefont{Sapeta and Wiedemann}(2008)}]{Sapeta:2008aa}
\bibinfo{author}{\bibfnamefont{S.}~\bibnamefont{Sapeta}} \bibnamefont{and}
  \bibinfo{author}{\bibfnamefont{U.~A.} \bibnamefont{Wiedemann}},
  \bibinfo{journal}{Eur. Phys. J.} \textbf{\bibinfo{volume}{C55}},
  \bibinfo{pages}{293} (\bibinfo{year}{2008}).

\bibitem[{\citenamefont{Liu and Fries}(2008)}]{Fries:2009aa}
\bibinfo{author}{\bibfnamefont{W.}~\bibnamefont{Liu}} \bibnamefont{and}
  \bibinfo{author}{\bibfnamefont{R.~J.} \bibnamefont{Fries}},
  \bibinfo{journal}{Phys. Rev. C} \textbf{\bibinfo{volume}{77}},
  \bibinfo{pages}{054902} (\bibinfo{year}{2008}).

\end{thebibliography}
%

\end{document}